\documentclass[letterpaper]{article} 
\usepackage[draft]{aaai2026}  
\usepackage{times}  
\usepackage{helvet}  
\usepackage{courier}  
\usepackage[hyphens]{url}  
\usepackage{graphicx} 
\usepackage{natbib}  
\usepackage{caption} 
\usepackage{algorithm}
\usepackage{algorithmic}
\newtheorem{remark}{Remark}
\usepackage{pdfpages}
\usepackage{newfloat}
\usepackage{listings}
\DeclareCaptionStyle{ruled}{labelfont=normalfont,labelsep=colon,strut=off} 
\floatstyle{ruled}
\newfloat{listing}{tb}{lst}{}
\floatname{listing}{Listing}
\usepackage{booktabs}
\usepackage{multirow}
\usepackage{amsmath}
\usepackage{amsfonts}

\title{Multi-Granularity Adaptive Time-Frequency Attention Framework for Audio \\Deepfake Detection under Real-World Communication Degradations}
\author {
    Haohan Shi\textsuperscript{\rm 1},
    Xiyu Shi\textsuperscript{\rm 1},
    Safak Dogan\textsuperscript{\rm 1},
    Tianjin Huang\textsuperscript{\rm 2},
    Yunxiao Zhang\textsuperscript{\rm 2}
}
\affiliations {
    \textsuperscript{\rm 1}Institute for Digital Technologies, Loughborough University London\\
    \textsuperscript{\rm 2}Department of Computer Science, University of Exeter\\
    h.shi@lboro.ac.uk, x.shi@lboro.ac.uk, s.dogan@lboro.ac.uk, t.huang2@exeter.ac.uk, y.zhang12@exeter.ac.uk
}

\usepackage{bibentry}

\begin{document}

\maketitle

\begin{abstract}
The rise of highly convincing synthetic speech poses a growing threat to audio communications. Although existing Audio Deepfake Detection (ADD) methods have demonstrated good performance under clean conditions, their effectiveness drops significantly under degradations such as packet losses and speech codec compression in real-world communication environments. In this work, we propose the first unified framework for robust ADD under such degradations, which is designed to effectively accommodate multiple types of Time-Frequency (TF) representations. The core of our framework is a novel Multi-Granularity Adaptive Attention (MGAA) architecture, which employs a set of customizable multi-scale attention heads to capture both global and local receptive fields across varying TF granularities. A novel adaptive fusion mechanism subsequently adjusts and fuses these attention branches based on the saliency of TF regions, allowing the model to dynamically reallocate its focus according to the characteristics of the degradation. This enables the effective localization and amplification of subtle forgery traces. Extensive experiments demonstrate that the proposed framework consistently outperforms state-of-the-art baselines across various real-world communication degradation scenarios, including six speech codecs and five levels of packet losses. In addition, comparative analysis reveals that the MGAA-enhanced features significantly improve separability between real and fake audio classes and sharpen decision boundaries. These results highlight the robustness and practical deployment potential of our framework in real-world communication environments. 
\end{abstract}


\section{Introduction}\label{section1}
The rapid advancement and widespread adoption of speech synthesis technologies have enabled the imitated human voices to be more convincing \cite{bisogni2024acoustic}. 
This has raised serious concerns about the potential misuse of deepfake audio in the real world, including identity impersonation \cite{news1, news3}, phone scams \cite{2020fraudsters}, mis/disinformation spread \cite{news4}, and unauthorized bank access \cite{news2}.
To address the growing threats, several international competitions, such as ASVspoof \cite{asvspoof2019,asvspoof2021,wang2024asvspoof} and the Audio Deep Synthesis Detection Challenge \cite{add2022}, have been launched to promote the development of standardized evaluation protocols and detection methods.
Consequently, Audio Deepfake Detection (ADD) has emerged as a critical research area in speech and security communities.

Recent studies have achieved notable progress in ADD under clean conditions, where audio inputs are high-fidelity and unaffected by communication systems. 
However, 
many of them neglect the impact of real-world communication degradations \cite{haohan, cohen2022study,besacier2003overview}, creating a substantial gap between experimental settings and practical communication scenarios, and often causing a severe performance degradation.

\begin{figure}[ht]
  \centering
  \includegraphics[width=0.5\textwidth]{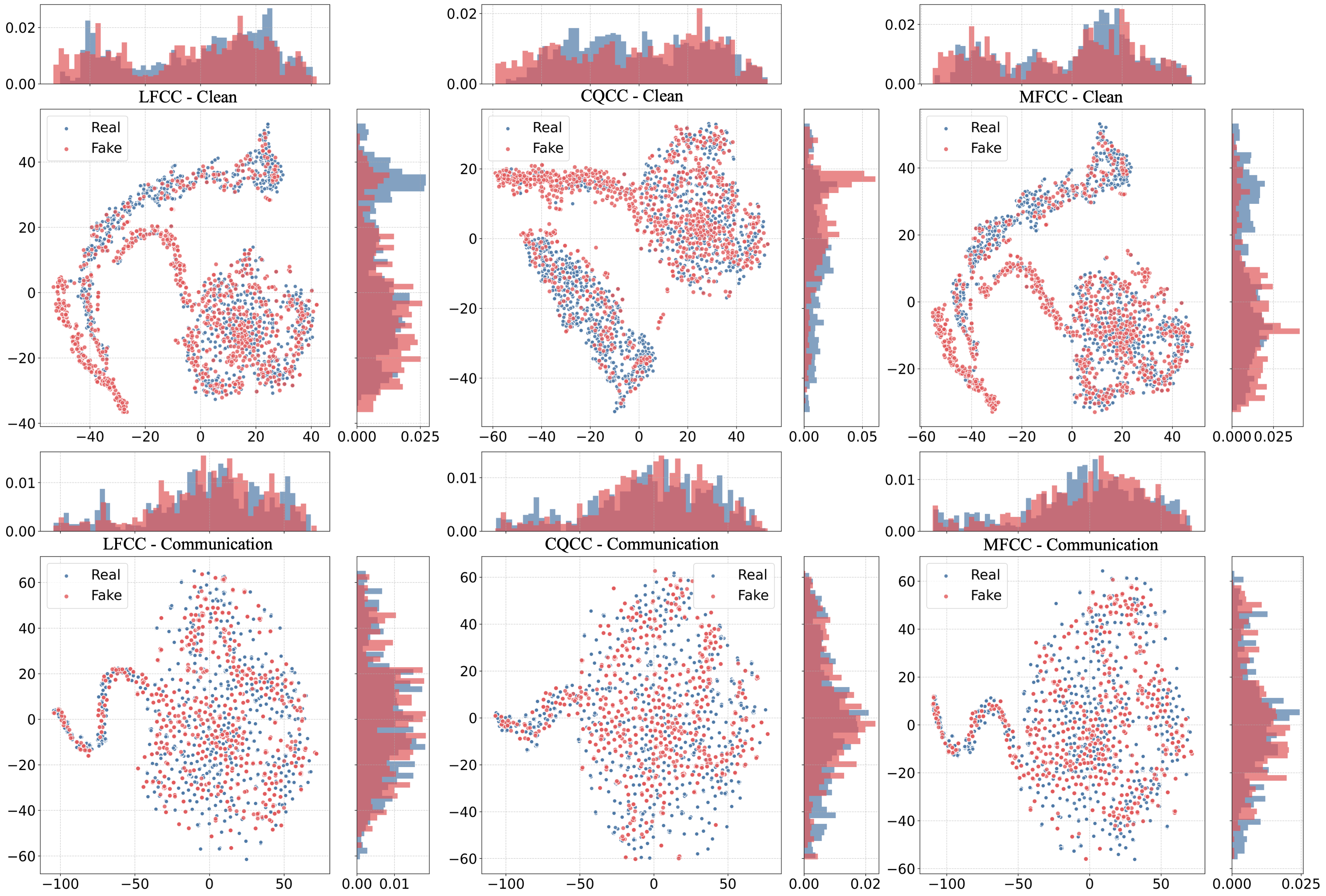}
  \caption{t-SNE \cite{tsne} visualizations of real and fake audio samples without (i.e., Clean) and with communication degraded effects (i.e., Communication) across different TF representations. The degradations are primarily due to speech codec compression and packet losses. Marginal histograms indicate sample density patterns along each axis. See more details in Appendix \ref{fig1details}.}
  \label{fig1}
\end{figure}

In real-world applications (e.g.,  video conferencing, Voice over Internet Protocol/Voice over Long Term Evolution calls, and broadcasting),
audio signals are rarely transmitted or received without quality degradation \cite{besacier2003overview,voip, volte, molisch2012wireless, todisco2017constant}. Instead, they are often corrupted by lossy compression, network congestion, and other transmission artifacts \cite{cohen2022study}. 
In particular, Figure \ref{fig1} illustrates the projection of high-dimensional Time-Frequency (TF) representations of real and deepfake audio samples, both with and without communication degraded effects, into a 2D space using t-SNE for visual analysis. We include Linear-Frequency Cepstral Coefficients (LFCC), Constant-Q Cepstral Coefficients (CQCC), and Mel-frequency Cepstral Coefficients (MFCC). 
We notice that
communication degradation causes more dispersed feature distributions and blurrier class boundaries (bottom three images), thereby significantly increasing the difficulty of ADD compared to clean conditions (top three images).
Such a difference highlights the need for the communication-aware ADD design and deployment in real-world applications.  

In this paper, for the first time, we propose a unified framework to address the challenges of ADD under real-world communication degradations, considering both varying Packet Loss Rates (PLR) and speech codec compression. The framework is evaluated using three widely adopted TF representations, LFCC, CQCC and MFCC.
Specifically, to deal with the diverse types of real-world communication degradations, our framework incorporates multiscale global and local receptive fields, which allow the simultaneous extraction of multi-granularity features from TF representations.
Furthermore, a novel adaptive fusion mechanism is introduced to dynamically adjust the attention focus based on the quality and characteristics of the degraded input audio. This design allows our framework to effectively localize and amplify subtle forgery traces across different real-world degradations, while maintaining high feature separability and clear decision boundaries even after severe communication transmission distortions.
We evaluated ADD detection performance with our proposed framework and state-of-the-art (SOTA) methods as baseline across six speech codec types and five PLR levels, resulting in 30 real-world communication degradation scenarios. The results show that our framework consistently outperforms the SOTA baselines under diverse degradation conditions.

Our main contributions are as follows:
\begin{itemize}
\item To the best of our knowledge, we are the first to propose a unified framework specifically targeting audio deepfake detection under diverse real-world communication degradations.

\item Our framework dynamically emphasizes salient time-frequency regions via multi-granularity attention and adaptive fusion, ensuring robust, effective, and efficient detection across various real-world communication degradations, highlighting its potential for practical deployment.

\item Our framework outperforms SOTA baselines under clean conditions and 30 types of real-world communication degradations (spanning six speech codecs and five PLR levels), and it significantly enhances feature separability and decision boundary clarity.

\end{itemize}





\section{Related Work}\label{section2}

\paragraph{Audio deepfake detection.}

Recent speech synthesis methods have greatly lowered the barrier to generating high-quality fake audio \cite{van2016wavenet,shi2024,shen2018natural,ren2020fastspeech,kumar2019melgan,yamamoto2020parallel,kong2020hifi, kong2020diffwave}, making ADD urgently needed.
Early studies on ADD primarily focused on traditional machine learning approaches, which relied on the combination of handcrafted acoustic features and classifiers, such as Gaussian Mixture Models and Support Vector Machines \cite{ref15,ref19,ref37, ref35}.
With the development of deep learning, various architectures including Convolutional Neural Networks, Deep Neural Networks, Long Short-Term Memory, and attention mechanisms have been introduced \cite{ref2, ref24, ref29, ref7, ref49}. These approaches learn discriminative features from raw waveforms or TF representations, significantly improving detection accuracy.
Recently, Self-Supervised Learning methods such as Wav2Vec \cite{ref22, ref23, ref39}, WavLM \cite{ref32}, and XLS-R \cite{ref15} have been adopted to reduce reliance on labelled data. However, they often require more computational resources.
In addition, novel physiological-based features have been proposed to capture human-specific characteristics, such as breathing-talking-silence \cite{ref27}, human vocal tract \cite{ref28} and linguistic styles \cite{zhu2024slim}.

\paragraph{Toward real-world communication degradation.}









In real-world communication scenarios, lossy transmission channels introduce a range of distortions, such as packet losses, bandwidth constraints, jitter, and codecs compression.
Although the recent ASVspoof5 challenge \cite{wang2024asvspoof} attempted to incorporate codec-induced distortions using AMR \cite{amrwb}, Speex \cite{speex}, and Opus \cite{opus} in its evaluation dataset, the approach remains limited and lacks systematic consideration. The selected codecs are outdated, while widely adopted modern codecs such as EVS \cite{evs} and IVAS \cite{IVAS}, which are standards in current 4G/5G mobile communication, are not covered.
Recent studies \cite{shim2023construct, sahidullah2025shortcut, shih2024does, chettri2023clever} expose shortcut learning and over-reliance on artifacts in ADD models under clean conditions, but largely ignore real-world degradations such as codec compression or packet loss. AASIST3 \cite{borodin2024aasist3} improves generalization via self-supervised learning but does not address transmission-induced distortions.
More importantly, these efforts fail to simulate the real-world communication degradations and do not provide insights into how different levels of lossy transmission quality affect the ADD methods.

A recent study \cite{haohan} first highlighted the impact of real-world communication degradations on ADD by introducing the ADD-C test dataset and an augmentation strategy, revealing that models trained on clean data suffer substantial performance drops under degraded communication scenarios.
However, this work primarily focused on dataset construction and data augmentation, leaving open the challenge of designing detection architectures that are both robust against real-world communication degradations and sensitive to forgery patterns.
Building on these insights, our work is the first to propose a unified framework that enables robust and generalizable ADD performance across diverse real-world communication degradations. The proposed framework outperforms SOTA baselines and significantly enhances feature separability and decision boundary clarity, which is an essential step towards practical real-world deployment.

\section{Methodology}\label{section3}

\subsection{Motivation and Communication Awareness}

In real-world communication scenarios, audio signals suffer from both lossy codec compression and random packet loss, introducing structured and stochastic distortions across the time and frequency domains. These distortions can significantly impact audio quality, masking or erasing the features used by detection methods to identify manipulated audio, resulting in a substantial performance drop for existing ADD methods \cite{haohan,cohen2022study, besacier2003overview,molisch2012wireless, todisco2017constant}.

To address this challenge, we design a communication-aware framework that explicitly models multi-scale, location-sensitive, and dynamically adaptive feature reliability. Our architecture is inspired by prior work in robust audio classification and spoofing detection \cite{lavrentyeva2019stc, valenti2017convolutional}, and we introduce the core component: Multi-Granularity Adaptive Time-Frequency Attention (MGAA), drawing insights from \cite{lin2017feature, wang2018non} to capture both global context and fine-grained distortions.
Our MGAA comprises three sub-modules:
\begin{itemize}
\item Global Time-Frequency Attention (GTFA): Inspired by Squeeze-and-Excitation networks \cite{hu2018squeeze} and temporal-frequency attention \cite{yadav2020frequency}, we use GTFA to capture global temporal-frequency dependencies, helping mitigate global distortions such as spectral flattening and temporal smearing.

\item Local Time-Frequency Attention (LTFA): Inspired by CBAM \cite{woo2018cbam}, our LTFA uses localized receptive fields to focus on spatially confined corruptions like packet loss or codec-induced artifacts.

\item Adaptive Fusion Module (AFM): Inspired by dynamic fusion techniques \cite{jia2016dynamic, li2019selective}, we use AFM to enable content-aware weighting of multiple attention pathways, allowing the model to adaptively emphasize relevant features based on degradation characteristics.
\end{itemize}

Overall, the proposed framework is inherently communication-aware, effectively capturing discriminative features under real-world communication degradations and addressing various distortion types. This design ensures robust and generalizable spoofing detection across diverse communication conditions.

\subsection{Framework overview}
\begin{figure}[t]
  \centering
  \includegraphics[width=0.48\textwidth]{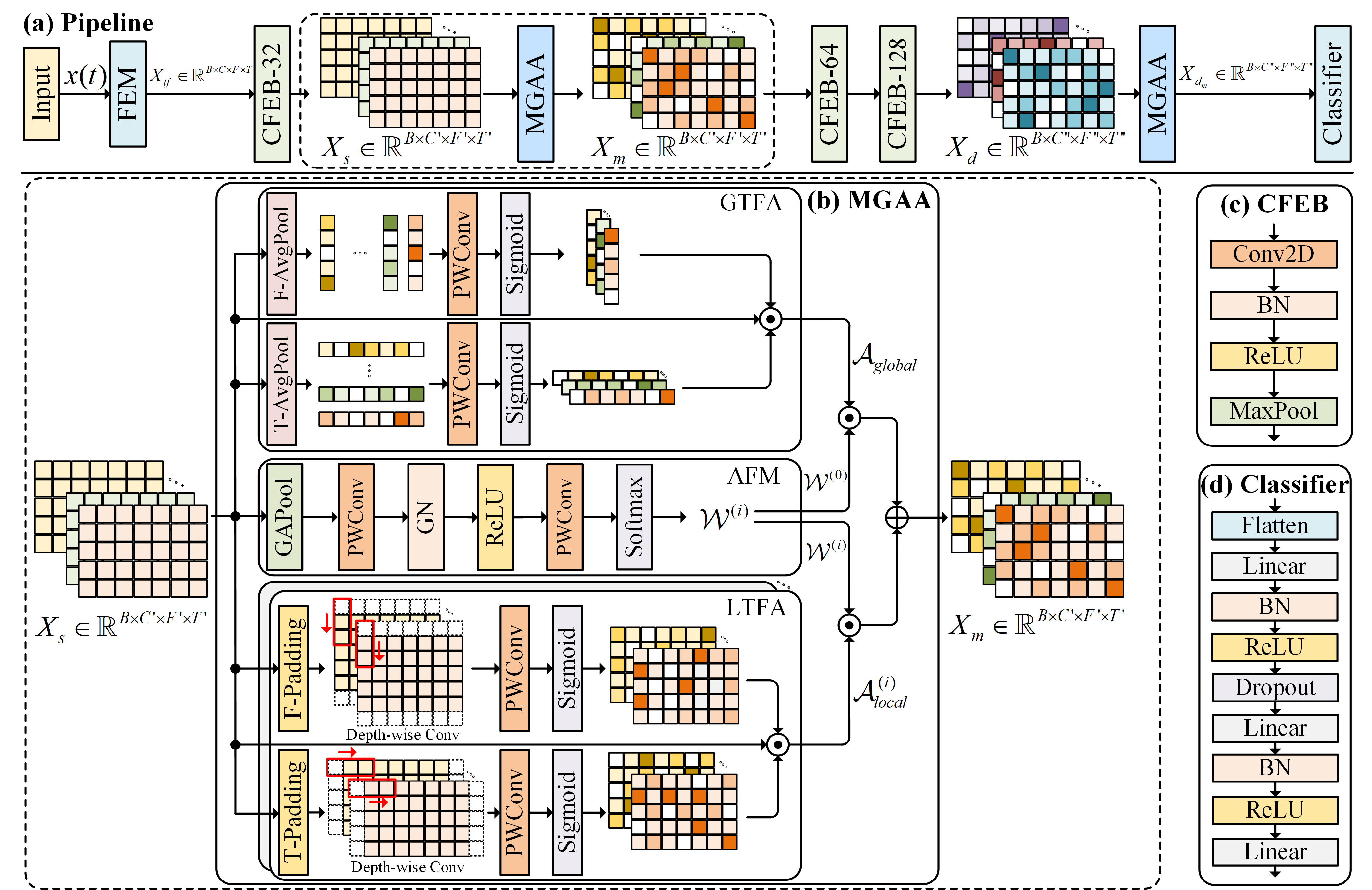}
  \caption{Proposed framework. (a) The processing pipeline; (b) Multi-Granularity Adaptive Time-Frequency Attention; (c) Convolutional Feature Embedding Blocks; (d) The Classifier.}
  \label{arch}
\end{figure}

The architecture of the proposed framework is shown in Figure \ref{arch}. 
Let \(x(t)\in \mathbb{R}^{1\times S}\) denote the input audio signal in the time domain with length \(S\), and let the binary ground-truth label \(y \in \{0,1\}\) indicate whether the audio is real (\(y=0\)) or fake (\(y=1\)). The objective is to learn a discriminative function \(f_\theta(X) \to y\), where \(\theta\) represents all trainable parameters in the framework, and \(X\) is the TF features.
The input audio signal \(x(t)\) is firstly processed by the Feature Extraction Module (FEM), which computes the corresponding TF representation. We denote the output of the FEM as \(X_{tf} \in \mathbb{R}^{B\times C\times F \times T}\), where \(B\) is the batch size, \(C\) is the number of feature channels, \(F\) and \(T\) represent the frequency and temporal dimensions, respectively.
The extracted TF features \(X_{tf}\) are then passed into the first shallow Convolutional Feature Embedding Blocks (CFEB-32), which extract shallow-level embedding features. Let the output of the CFEB-32 be \(X_{s} \in \mathbb{R}^{B\times C'\times F' \times T'}\), where \(B'\), \(C'\) and \(F'\) represent the updated dimensions.
\(X_{s}\) is then processed by the proposed MGAA and the output is denoted as \(X_{m}\in \mathbb{R}^{B\times C'\times F' \times T'}\).
To capture deep-level features, \(X_{m}\) is further processed by CFEB-64 and CFEB-128, which progressively increase the receptive field and feature depth. The output from CFEB-128 is denoted as \(X_{d}\in \mathbb{R}^{B\times C''\times F'' \times T''}\), where \(B''\), \(C''\) and \(F''\) represent the new updated dimensions after deep-level feature embedding. \(X_{d}\) is then passed through a second-stage MGAA to repeat the process at a deeper level to form the final encoded feature, denoted as \(X_{d_m}\).
Finally, \(X_{d_{m}}\) is flattened and passed through a fully connected Classifier to output the resulting binary prediction.

\subsection{Framework components}

\paragraph{FEM.}
The design of the FEM was motivated by the feature extraction process outlined in \cite{asvspoof2021}.
Each TF feature is computed on fixed-length 4s audio segments and follows a unified extraction pipeline comprising spectral decomposition, filterbank projection, logarithmic scaling, and Discrete Cosine Transform (DCT). By denoting the static cepstral coefficients from the TF representation as \(X(j,n)\in \mathbb{R}^{F\times T}\), where \(j\in\{1,\cdots,F\}\) and \(n\in\{1,\cdots,T\}\), we get:
\begin{align*}
    X(j, n) &= \sum_{i=0}^{M-1} \log ( \sum_{k} H_i^{(\mathcal{F})}(f_k) \cdot |\mathcal{T}\{x(t)\}(k, n)|^2 + \varepsilon ) \cdot \nonumber \\ 
    &\cos( \frac{\pi j(i + 0.5)}{M} ),
\end{align*}
where \( \mathcal{T} \{\cdot\} \in \{ \text{STFT}, \text{CQT} \} \) represents the operator of Short Time Fourier Transform and Constant-Q Transform with hop lengths set to 512, \(\varepsilon=1e\scriptstyle^{-10}\) ensures numerical stability, 
\(\textstyle{f_k}\) is the center frequency of the \(k\)-th frequency bin, and \smash{\(H\scriptstyle_i^{(\mathcal{F})}\textstyle(f_k)\)} is the \(i\)-th filter under frequency scale \(\mathcal{F} \in \{ \text{linear},\text{log-scale},\text{mel} \} \), corresponding to LFCC, CQCC and MFCC, respectively. The output of the DCT for each TF feature yields 20 static cepstral coefficients per frame.
To capture short-term temporal dynamics in cepstral trajectories, we further compute the first and second-order derivatives using a regression window of \(R=4\), defined as:
\begin{align*}
    &\Delta X(j, n) = \frac{\sum_{r=1}^R r \cdot (X(j, n + r) - X(j, n - r))}{2 \sum_{r=1}^R r^2}, \quad \nonumber  \\ 
    &\Delta^2 X(j, n) = \Delta (\Delta X(j, n)),
\end{align*}
where \(\Delta\) is the derivative operator. This results in three components of cepstral features, and the final TF representation is \(X_{tf}=[X,\Delta X,\Delta^{2} X] \in \mathbb{R}^{C\times F\times T}\), where \(C=1\), \(F=60\) and \(T=126\).

\paragraph{CFEB.}
The CFEB was designed for feature embedding. 
When we denote \(\lambda\in \mathbb{R}^{B\times C_{in}\times F \times T}\) as the input to the CFEB, the output is computed as:
\begin{align*}
    \text{CFEB-}C_{out}(\lambda)= \mathcal{P}(\sigma_{r}(\mathcal{N}_b(\mathcal{F}(\lambda)))) \in \mathbb{R}^{B \times C_{out} \times \frac{F}{2} \times \frac{T}{2}},
\end{align*}
where \( C_{out} \in\{32,64,128\}\), \(\mathcal{F}(\cdot)\) is the convolution operation with kernel size three and padding one, \(\mathcal{N}_b(\cdot)\) the batch normalization, \(\sigma_{r}(\cdot)\) the ReLU, and \(\mathcal{P}(\cdot)\) the max pooling with stride two.

\paragraph{MGAA.}
Let's denote the input of the MGAA as \(\xi \in \mathbb{R}^{B \times C \times F \times T}\). The GTFA captures long-range global dependencies across time and frequency dimensions. The output is computed as:
\begin{align*}
    \mathcal{A}_{\text{global}}(\xi) &= \xi \odot (\sigma_s(V_f *\mathcal{P}_{avg_{f}}(\xi))) \odot \nonumber \\
    &(\sigma_s(V_t * \mathcal{P}_{avg_{t}}(\xi)))  \in \mathbb{R}^{B \times C \times F \times T},
\end{align*}
where \(\mathcal{P}_{avg_f}(\cdot)\) and \(\mathcal{P}_{avg_t}(\cdot)\) represent the adaptive average pooling over the frequency and time dimension, \(V_f\) and \(V_t\) the pointwise convolution to facilitate channel-wise interaction and learnable weighting across each spatial location, \(\sigma_s\) the sigmoid activation, \(*\) the convolution operator and \(\odot\) the element-wise product.

The LTFA focuses on capturing fine-grained and localized patterns via multiple attention branches with different window sizes \(k_i \in \{k_1, k_2, \dots, k_n\}\), where \(n\) is the number of local attention branches. If we define the input of each branch with window size \(k_i\), the outputs of LTFA are computed as:
\begin{align*}
    \mathcal{A}^{(i)}_{\text{local}}(\xi) &= \xi \odot (\sigma_s(V^{(i)}_f * \mathcal{F}^{(i)}_{DW_f}(\xi))) \odot  \nonumber \\
    &(\sigma_s(V^{(i)}_t * \mathcal{F}^{(i)}_{DW_t}(\xi)))  \in \mathbb{R}^{B \times C \times F \times T},
\end{align*}
where \smash{\(\mathcal{F}\scriptstyle^{(i)}_{DW_f}\textstyle(\cdot)\)} represents a depth-wise convolution with kernel size (\(k_i, 1\)) and appropriate padding along the frequency dimension, capturing vertical local features; \smash{\(\mathcal{F}\scriptstyle^{(i)}_{DW_t}\textstyle(\cdot)\)} represents a depth-wise convolution with kernel size (\(1, k_i\)) and appropriate padding along the time dimension, capturing horizontal local features;
\smash{\(V\scriptstyle^{(i)}_f\)} and \smash{\(V\scriptstyle^{(i)}_t\)} represent pointwise convolutions. 
The \smash{\(\mathcal{F}\scriptstyle^{(i)}_{DW_f}\textstyle(\cdot)\)} and \smash{\(\mathcal{F}\scriptstyle^{(i)}_{DW_t}\textstyle(\cdot)\)} efficiently capture multiple local dependencies within fixed-size windows, allowing the focus on relevant local time-frequency patterns. 
The subsequent \smash{\(V\scriptstyle^{(i)}_f\)} and \smash{\(V\scriptstyle^{(i)}_t\)} enable inter-channel information exchange, while the \(\sigma_s\) generates attention maps in the range of \([0,1]\), which highlight important features when applied multiplicatively to the input.

The AFM dynamically adjusts the contribution of different attention branches based on the input feature map. The weight of each branch is computed as \(\mathcal{W}(\xi) = \sigma_{sf}(V_g * \sigma_r(\mathcal{N}_g(V_r * \mathcal{P}_{g}(\xi))\), where \(\mathcal{P}_{g}(\cdot)\) represents global average pooling to capture a channel-wise summary of the entire feature map while ensure efficiency, \(\sigma_{sf}\) is softmax activation, \(V_r\) is dimensionality reduction pointwise convolution, \(\mathcal{N}_g(\cdot)\) is group normalization, \(V_g\) is pointwise convolution to generate weights for each attention branch.
The final output of MGAA is computed as:
\begin{align*}
    &\textstyle MGAA_{out}(\xi) = \sum_{i=0}^{n} \mathcal{W}^{(i)}(\xi) \cdot \mathcal{A}^{(i)}(\xi),
\end{align*}
where $\sum_{i=0}^{n} \mathcal{W}^{(i)}(\xi) = 1$, $\mathcal{W}^{(i)}(\xi)\geq 0$  for all $i$, and \(\mathcal{A}^{(0)}(\xi)=\mathcal{A}_{\text{global}}(\xi)\).
This enables the framework to dynamically adjust the contribution of each attention branch based on the input characteristics, emphasize the most relevant granularity of features for each specific input sample, and combine global with multiple local patterns to enhance feature representation.

\paragraph{Classifier.}
The final TF feature embeddings are flattened and passed to a three-layer fully connected neural network for classification.
Let's denote the flattened input as \(F\in \mathbb{R}^{B\times d}\), where \(d\) is the dimension of the flattened features. The output of the classifier is:
\begin{align*}
    \mathcal{C}_{out}(F) = W_3 \cdot \sigma_r(\mathcal{N}_b(W_2 \cdot \mathrm{D}(\sigma_r(\mathcal{N}_b(W_1 \cdot F))))) \in \mathbb{R}^{B \times 2},
\end{align*}
where \(W_1 \in \mathbb{R}^{256 \times d}\), \(W_2 \in \mathbb{R}^{ 64\times 256}\) and \(W_3\in \mathbb{R}^{2 \times 64}\) represent the weight matrix, \(\mathrm{D}(\cdot)\) is the dropout (0.3). The proposed framework is trained using cross-entropy loss:
\begin{align*}
    \mathcal{L}(\theta) = - \mathbb{E}_{(X, y) \sim \mathcal{D}} \left[ y \log f_\theta(X) + (1 - y) \log(1 - f_\theta(X)) \right].
\end{align*}

\section{Experiments}\label{section4}
\subsection{Setup}\label{section4.1}

\paragraph{Dataset, training and evaluation.}
Six publicly available speech datasets, \textit{Fake-or-Real (FoR)} \cite{FoR}, \textit{Wavefake} \cite{wavefake}, \textit{LJSpeech} \cite{ljspeech}, \textit{MLAAD-EN} \cite{mlaad}, \textit{M-AILABS} \cite{mailabs}, and \textit{ASVspoof 2021 Logical Access (ASVLA)} \cite{asvspoof2021}, are selected for dataset construction.
We combine all the real and fake utterances across these datasets to form a new comprehensive dataset, denoted as \(\mathcal{D}\). Such a new dataset comprises 130,041 real and 240,373 fake utterances with 36 audio deepfake algorithms in total.
We further process dataset \(\mathcal{D}\) using the data augmentation strategy proposed in \cite{haohan}. It results in an expanded dataset \(\mathcal{D}_{com}\), which includes 640,205 real and 1,191,865 fake utterances, covering 30 types of real-world communication degradations. 
These degradations are generated in a balanced manner using six speech codecs, i.e., \textit{AMR-WB} \cite{amrwb}, \textit{EVS} \cite{evs}, \textit{IVAS} \cite{IVAS}, \textit{Speex(WB)} \cite{speex}, \textit{SILK} \cite{silk}, and \textit{OPUS} \cite{opus}. Each of them is applied under five different PLRs (0\%, 1\%, 5\%, 10\% and 20\%). Further details are available in Appendix \ref{dataset}.

Dataset \(\mathcal{D}_{com}\) is split into 80\% for training and 20\% for validation. We select a batch size of 256 and five epochs, while early stopping \cite{earlystop} with a patience of three is applied to prevent overfitting. 
AdamW optimizer \cite{adamW} is employed for weight updates, and a cosine annealing \cite{cos} is adopted to dynamically adjust the learning rate throughout training.

For evaluation of the ADD methods, we use the ADD-C test dataset in \cite{haohan}. It comprises six conditions \(C_0\) to \(C_5\).
Specifically, \(C_0\) corresponds to the clean, undistorted condition. \(C_1\) to \(C_5\) represent five progressively severe degradation levels, which incorporate six different speech codecs to simulate codec compression, with PLR of 0\%, 1\%, 5\%, 10\%, and 20\%, respectively.
These conditions simulate real-world communication degradations, where both codec-induced compression artifacts and channel transmission-induced packet losses jointly affect the 
audio
quality (Further details are in Appendix \ref{addcdetails}).

\paragraph{Evaluation metrics.}
Equal Error Rate (EER) is chosen as the evaluation metric for assessing ADD methods \cite{asvspoof2019,asvspoof2021,wang2024asvspoof}. EER represents the error rate when the false acceptance rate and false rejection rate of the ADD method are equal, offering a single, intuitive measure that effectively balances both types of errors. Lower EER suggests better performance.

\paragraph{Baselines.}
To ensure a rigorous comparative evaluation, we implement and evaluate ten SOTA baselines under identical training and evaluation conditions. These include \textit{GMM-CQCC} \cite{asvspoof2021}, G\textit{MM-LFCC} \cite{asvspoof2021}, \textit{LCNN} \cite{asvspoof2021}, \textit{RawNet2} \cite{rawnet2}, \textit{RawGAT-ST} \cite{rawgat}, \textit{AASIST} \cite{aasist}, \textit{AASIST-L}\cite{aasist}, \textit{FC-LFCC} \cite{haohan}, \textit{FC-CQCC} \cite{haohan}, and \textit{FC-waveform} \cite{haohan}.


All models are trained on a PC equipped with an Intel Core i7-12700K CPU and an NVIDIA RTX 3090 GPU (24GB RAM) using the training dataset \(\mathcal{D}_{com}\), and evaluated on the ADD-C dataset to ensure a fair comparison across all baselines. The hyperparameters are set according to the configuration specification in the referenced literature.

\subsection{Experimental results}\label{section4.2}

\paragraph{Detection performance and computational complexity.}
Table \ref{tab1} presents the detection performance and computational complexity comparison between ten baselines and the proposed framework.
Using MFCC as input, our framework achieves the lowest average EER of 0.15\%.
Although performance slightly decreases under \(C_5\), our framework still outperforms all baselines across all degradation conditions.
In addition, LFCC and CQCC inputs consistently yield low EER across all conditions, with average scores of 0.22\% and 0.67\%, respectively, both ranking in the top five among all comparison methods. This demonstrates the framework's generalization across diverse types of TF representations.

\begin{table}[htb]
\centering
\setlength{\tabcolsep}{0.5mm}
\scalebox{0.77}{
\begin{tabular}{ccccccccccc}
\toprule
\multirow{2}{*}{Model}  & \multicolumn{7}{c}{EER(\%) \(\downarrow\)}  & \multirow{2}{*}{\#Para.} & \multirow{2}{*}{Time}\\ \cmidrule(r){2-8}& \(C_{0}\)  & \(C_{1}\) & \(C_{2}\) & \(C_{3}\) & \(C_{4}\) & \(C_{5}\) & Avg. & (million) & (hours)\\
\midrule
GMM-CQCC & 45.15 & 44.99 & 44.37 & 44.27 & 44.17 & 43.98 & 44.38 &0.12 &5.43\\
GMM-LFCC & 46.87 & 47.51 & 47.45 & 47.37 & 47.10 & 46.68 & 47.17 &0.12&14.33\\
LCNN & 0.63 & 0.86 & 0.86 & 0.94 & 1.12 & 1.43 & 0.97 & 0.34 & 2.10\\
RawNet2 & 0.63 & 0.44 & 0.46 & 0.53 & 0.72 & 1.35 & 0.69 & 17.62 & 2.56\\
RawGAT-ST & 0.38 & 0.22 & 0.22 & 0.24 & 0.30 & 0.52 & 0.32 & 0.44 & 67.50\\
AASIST & 0.33 & 0.26 & 0.27 & 0.31 & 0.38 & 0.77 & 0.39 & 0.30 & 35.15\\
AASIST-L & 1.02 & 0.91 & 0.92 & 1.12 & 1.44 & 2.46 & 1.31 & 0.09 & 28.25\\
FC-CQCC & 33.35 & 33.58 & 33.58 & 33.47 & 33.44 & 33.64 & 33.51 &3.18 & 6.17\\
FC-LFCC & 1.20 & 1.44 & 1.48 & 1.61 & 1.82 & 2.85 & 1.73 & 15.35& 2.08\\
FC-wavform & 38.65 & 38.61 & 38.57 & 38.56 & 38.70 & 38.74 & 38.64 & 2.22&1.80\\
\midrule
Ours-LFCC & 0.11 & 0.12 & 0.12 & 0.14 & 0.22 & 0.61 & 0.22 & 3.74 & 2.17\\
Ours-CQCC & 0.31 & 0.36 & 0.36 & 0.44 & 0.71 & 1.84 & 0.67 & 3.74 &2.92\\
Ours-MFCC & \textbf{0.10} & \textbf{0.10} & \textbf{0.10} & \textbf{0.10} & \textbf{0.14} & \textbf{0.34} & \textbf{0.15} & 3.74 & \textbf{0.58}\\
\bottomrule
\end{tabular}}
\caption{Comparison of detection performance and computational complexity. \#Para. refers to the number of trainable parameters. Experiments were repeated three times with different random seeds, and average metric values are reported. Bold entries indicate the lowest value.}
\label{tab1}
\end{table}

\begin{remark}
We observe that baselines such as \textit{RawNet2}, \textit{RawGAT-ST}, \textit{AASIST}, and \textit{AASIST-L} perform worse under the clean condition \(C_0\) than under degraded conditions like \(C_1\). 
This is attributed to the absence of clean samples in \(\mathcal{D}_{com}\), which limits the baselines' generalization to clean audio. In contrast, our framework shows minimal performance variation between \(C_0\) and \(C_1\), highlighting strong robustness and cross-domain generalization ability without exposure to clean data during training.
Moreover, a consistent performance degradation is observed from 
\(C_1\) to \(C_5\) across nearly all methods, highlighting the impact of real-world communication degradations on ADD methods.
\end{remark}

\paragraph{Efficiency and practicality.}
To evaluate real-world deployment potential, we compare our framework with six top-performing methods, as shown in Table \ref{realtime}.
Our framework variants using MFCC and LFCC as input demonstrate the lowest GFLOPs (i.e., 0.13) and fastest inference times (i.e., 3.02 ms and 4.30 ms, respectively), while maintaining competitive or superior detection performance. These results reflect highly efficient model design and lightweight computational overhead, especially when compared to complex end-to-end models like \textit{RawGAT-ST} and \textit{AASIST}.
Notably, in terms of training efficiency, as shown in Table \ref{tab1}, our MFCC input model requires only 0.58 hours, achieving the best detection performance while being 116.38\(\times\) faster than \textit{RawGAT-ST} (i.e., 67.50 hours) and 60.60\(\times\) faster than \textit{AASIST} (i.e., 35.15 hours). 
This substantial reduction in training time illustrates the framework’s suitability for resource-constrained environments and rapid model deployment.
Overall, these results indicate the efficiency, practicality, and potential for real-world deployment.

\begin{table}[htb]
\centering
\setlength{\tabcolsep}{3mm}
\scalebox{0.8}{
\begin{tabular}{cccc}
\toprule
Model & GFLOPs & RTF (\%)& Infer time (ms)\\
\midrule
LCNN & 0.65 & 0.03 & 1.27 \\
RawNet2 & 1.55 & 0.12 & 4.73 \\
RawGAT-ST & 36.12 & 0.29 & 11.65\\
AASIST & 18.08 & 0.16 & 6.24\\
AASIST-L & 12.18 & 0.14 & 5.68\\
FC-LFCC & 0.18 & 0.14 & 5.30\\
\midrule
Ours-LFCC & 0.13 & 0.11 & 4.30\\
Ours-CQCC & 0.79 & 0.77 & 30.83\\
Ours-MFCC & 0.13 & 0.08 & 3.02\\
\bottomrule
\end{tabular}}
\caption{Comparison of practical efficiency. Giga Floating Point Operations Per Second (GFLOPs), Real-time Factor (RTF), and Infer time are reported, with results averaged over 100 runs.}
\label{realtime}
\end{table}

\paragraph{Cross-PLR and Cross-Codec Evaluation.} 

We evaluate the generalization ability of our framework under four challenging scenarios, as shown in Table \ref{cross}.
In the Unseen PLR setting, the framework trained on lower PLRs generalizes well to higher PLR (20\%) across three inputs. 
In the Unseen Codec scenario, it also performs competitively on codecs not seen during training, demonstrating strong codec robustness. 
The Unseen PLR+Codec excludes both a PLR and a codec, yields moderate degradation, but still maintains effective detection, especially with LFCC and CQCC. 
In the most challenging Unseen Severe setting, where both PLRs (10\%, 20\%) and codecs (IVAS, EVS, Speex) are excluded, the framework exhibits expected performance drops but still yields competitive results, particularly with MFCC.
These findings collectively validate the robustness and transferability of our framework under real-world deployment scenarios, where both the communication environment and codec configurations may vary unpredictably.

\begin{table}[ht]
\centering
\footnotesize
\setlength{\tabcolsep}{1mm}
\scalebox{0.75}{
\begin{tabular}{lccccccc}
\toprule
\multirow{2}{*}{Setting} & \multicolumn{2}{c}{PLR (\%)} & \multicolumn{2}{c}{Codec} & \multicolumn{3}{c}{Avg. EER (\%)} \\\cmidrule(lr){2-3} \cmidrule(lr){4-5} \cmidrule(lr){6-8}
 & Train & Test & Train & Test & LFCC & CQCC & MFCC \\
\midrule
Unseen PLR & 0,1,5,10 & 20 & All & All & 1.22 & 3.12 & 2.34 \\
Unseen Codec & All & All & A,O,S,I & E,Sp & 2.59 & 3.46 & 2.83 \\
Unseen PLR+Codec & 0,1,5,20 & 10 & A,O,S,E,Sp & I & 0.51 & 1.26 & 4.24 \\
Unseen Severe & 0,1,5 & 10,20 & A,O,S & I,E,Sp & 4.18 & 10.34 & 2.37 \\
\bottomrule
\end{tabular}}
\caption{Cross-condition evaluation on unseen PLRs and Codec. Codec abbreviations: A (\textit{AMRWB}), O (\textit{OPUS}), S (\textit{SILK}), E (\textit{EVS}), Sp (\textit{Speex}), I (\textit{IVAS}).}
\label{cross}
\end{table}

\begin{figure}[h]
  \centering
  \includegraphics[width=0.4\textwidth]{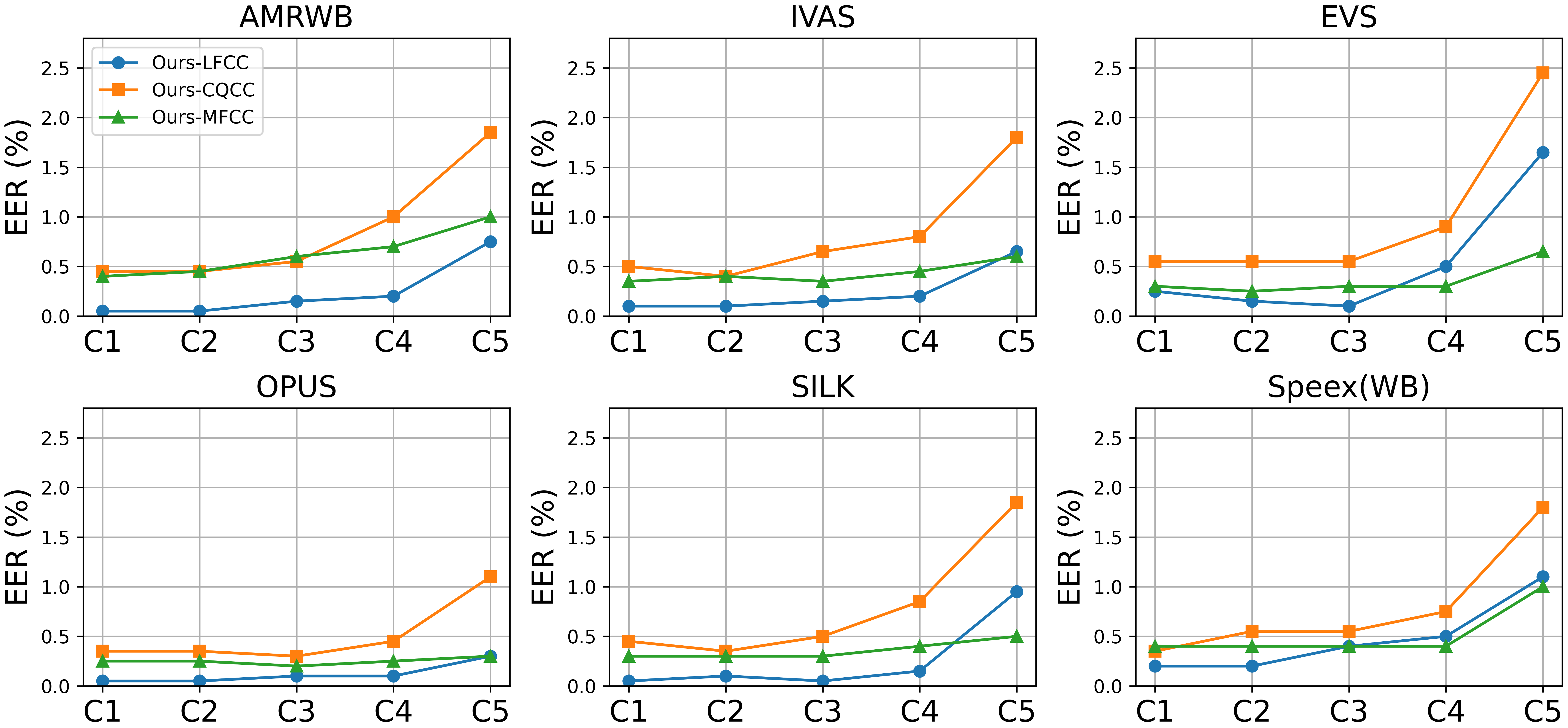}
  \caption{Comparison of detection performance with different speech codecs.}
  \label{fig:codec}
\end{figure}

\paragraph{Detection performance under different speech codecs.}\label{4.2.2}
Figure \ref{fig:codec} shows the codec-specific robustness of the proposed framework under \(C_1\) to \(C_5\) across three TF features.


\textit{OPUS} consistently delivers the most robust performance, achieving the lowest EER (0.29\%) with minimal variance, benefiting from its hybrid Linear Predictive Coding (LPC) and Constrained Energy Lapped Transform (CELT) architecture that preserves both temporal and frequency cues. \textit{SILK} and \textit{IVAS} also yield strong results (EER of 0.47\% and 0.50\%), with \textit{SILK+MFCC} showing particularly stable behavior.
In contrast, \textit{Speex(WB)} and \textit{EVS} exhibit the poorest performance (0.60\% and 0.63\%), with EER deteriorating under severe PLRs, suggesting that Code-Excited Linear Prediction (CELP)/Algebraic-CELP (ACELP) encoding and PLC mechanisms suppress anomalous TF patterns during quantization and packet loss concealment, thereby over-smoothing or removing subtle forgery artifacts.
\textit{AMR-WB} is based on ACELP and shows moderate robustness (0.58\%), performing well at mild PLRs but deteriorating from $C_4$.
t-SNE visualizations further support these findings: OPUS preserves clean class separability across all PLRs, while EVS and Speex suffer from collapsed distributions under severe degradation (see Appendix \ref{codectsneAnalysis} for detailed analysis).
These findings emphasize the important role of codec architecture in preserving or distorting the discriminative features used in ADD, offering valuable insights for future research on codec-aware or codec-agnostic detection systems. Overall, our framework demonstrates consistent and codec-resilient detection performance, underscoring its potential for real-world deployment and motivating further advances in robust ADD design.

\begin{figure}[h]
  \centering
  \includegraphics[width=0.45\textwidth]{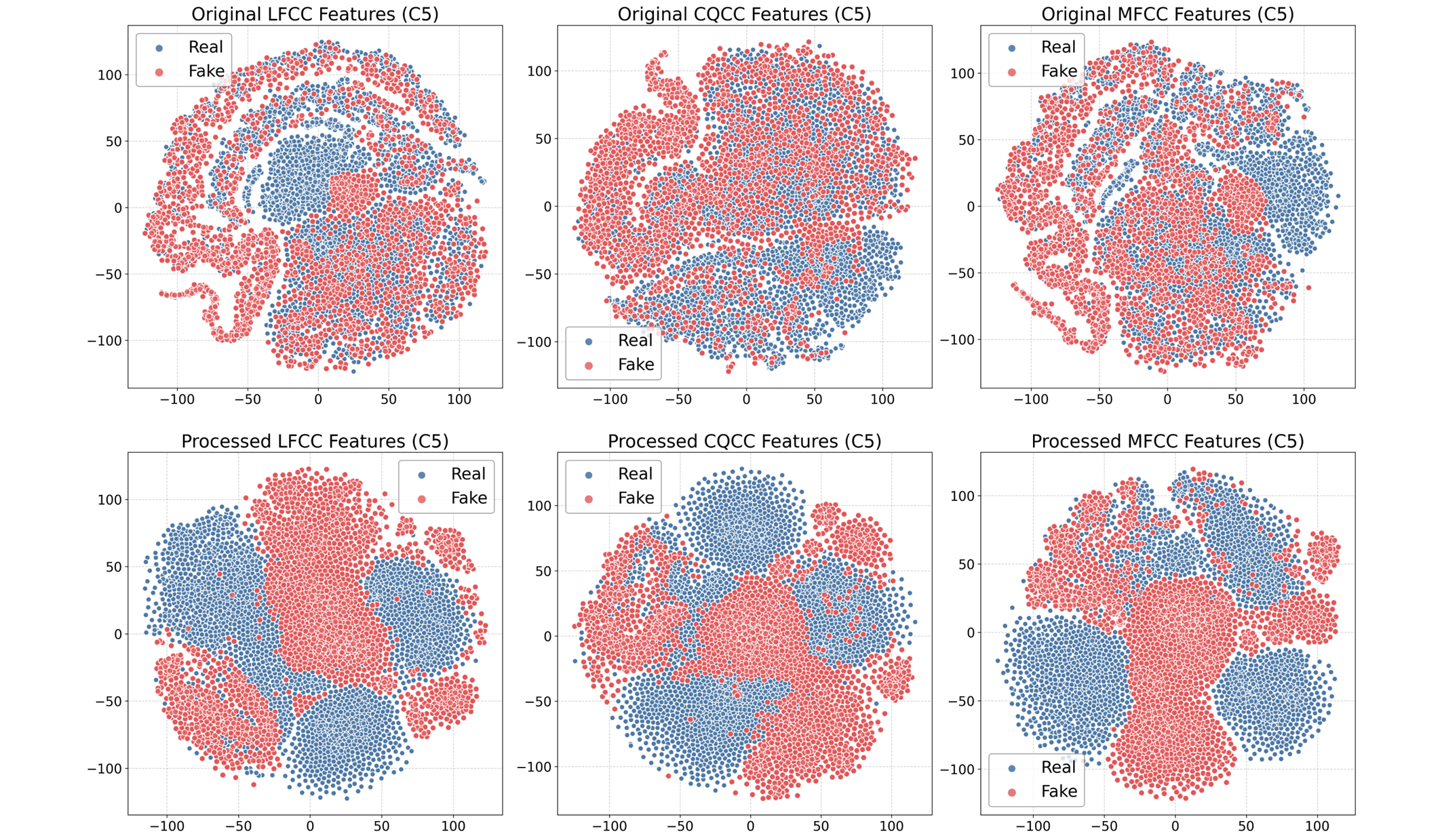}
  \caption{t-SNE visualizations of real and fake audio samples across different TF representations under \(C_5\). The top row represents the original features, and the bottom row represents the processed features extracted from the proposed framework before the Classifier.}
  \label{fig:tsne}
\end{figure}

\paragraph{Analysis of feature separability.}\label{tsnemain}
To qualitatively assess the representation learning capability of the proposed framework, we employ t-SNE to project the high-dimensional feature embeddings into a 2D space for visual analysis. This allows us to examine the separability of real and fake audio samples under \(C_0\) to \(C_5\).
Figure \ref{fig:tsne} shows the comparison across three types of TF representations under the most severe real-world communication degradations \(C_5\) (see Appendix \ref{tsne} for details of \(C_0\)-\(C_4\)).

As shown in the top rows of Figure \ref{fig:tsne} and Figure \ref{AppentsneC0}-\ref{AppentsneC4} in Appendix \ref{tsne}, the original TF features exhibit significant class entanglement. Real and fake samples are densely overlapped, indicating weak discriminative capacity when directly subjected to severe real-world communication degradations.
In contrast, the bottom rows illustrate that features extracted by our framework exhibit clearly separated and compact clusters for real and fake classes, with minimal inter-class confusion. 
This indicates that the framework effectively captures discriminative global and local patterns from corrupted signals, thus enhancing the downstream classification performance.
These visualizations qualitatively demonstrate that our framework substantially improves feature quality and class separability under real-world communication degradations.

\subsection{Ablation Studies}\label{section4.3}

\paragraph{Effects of different components.} 

Table \ref{ablation} presents the impact of removing or altering individual components of our framework. 
``Shallow'' and ``Deep'' indicate MGAA is placed only in shallow or deep layers, while (f) replaces AFM with a fixed equal-weight fusion of GTFA and LTFA.
Removing MGAA results in a notable decline in detection performance across all types of TF representations, confirming its critical role.
Using only GTFA or LTFA also degrades detection performance, though LTFA yields slightly better results, indicating that localized attention contributes more to capturing fine-grained forgery artifacts.
Placing MGAA in a shallow layer is consistently superior to deeper placement, indicating that early-layer attention can preserve more detailed TF representation.
Additionally, replacing AFM with a fixed fusion results in a significant performance drop, emphasizing the necessity of dynamically adapting the attention weights based on the different degraded inputs.

\begin{table}[ht]
\centering
\setlength{\tabcolsep}{1mm}
\scalebox{0.68}{
\begin{tabular}{cccccccccc}
\toprule
\multirow{2}{*}{Setting} & \multirow{2}{*}{GTFA} & \multirow{2}{*}{AFM} & \multirow{2}{*}{LTFA} & \multirow{2}{*}{Deep} & \multirow{2}{*}{Shallow} & \multirow{2}{*}{\(k\)} & \multicolumn{3}{c}{Avg. EER(\%) \(\downarrow\)} \\\cmidrule(r){8-10} &&&&&&&LFCC & CQCC & MFCC \\ \midrule   
(a) & \(\times\) & \(\times\) & \(\times\) & \(\times\) & \(\times\) & \multirow{6}{*}{\(\{3,5,7,9\}\)} & 0.83 & 1.87 & 1.36 \\
(b) & \(\bullet\) & \(\bullet\) & \(\times\) & \(\bullet\) & \(\bullet\) & & 0.57 & 1.61 & 1.04\\
(c) & \(\times\) & \(\bullet\) & \(\bullet\) & \(\bullet\) & \(\bullet\) & & 0.50 & 1.48 & 0.82\\
(d) & \(\bullet\) & \(\bullet\) & \(\bullet\) & \(\times\) & \(\bullet\) & & 0.51 & 1.31 & 0.65\\
(e) & \(\bullet\) & \(\bullet\) & \(\bullet\) & \(\bullet\) & \(\times\) & & 0.53 & 1.66 & 0.57 \\
(f) & \(\bullet\) & \(\times\) & \(\bullet\) & \(\bullet\) & \(\bullet\) & & 0.46 & 0.96 & 0.54 \\
\midrule
(g) & \(\bullet\) & \(\bullet\) & \(\bullet\) & \(\bullet\) & \(\bullet\) & \(\{3, 5\}\) & 0.58 & 0.92 & 0.62 \\
(h) & \(\bullet\) & \(\bullet\) & \(\bullet\) & \(\bullet\) & \(\bullet\) & \(\{3,5,7\}\) & 0.50 & 0.89 & 0.50 \\
(i) & \(\bullet\) & \(\bullet\) & \(\bullet\) & \(\bullet\) & \(\bullet\) & \(\{3,5,7,9,11\}\) & 0.67 & 0.98 & 0.59 \\
(j) & \(\bullet\) & \(\bullet\) & \(\bullet\) & \(\bullet\) & \(\bullet\) & \(\{3,5,7,9\}\) & \textbf{0.30} & \textbf{0.73} & \textbf{0.41} \\
\bottomrule
\end{tabular}}
\caption{Ablation studies of framework components and granularity configurations \(k\), where \(\bullet\) and \(\times\) denote inclusion and exclusion, respectively.}
\label{ablation}
\end{table}

\paragraph{Selection of granularity configurations.}

Window sizes \(k\) are selected based on their dual acoustic significance. In the time domain, with each value representing 31.75ms, our configurations correspond to specific linguistic units: \(k=3\) (i.e., 95ms) captures phoneme-level events, \(k=5\) (i.e., 159ms) spans formant transitions, \(k=7\) (i.e., 222ms) encompasses syllabic structures, and \(k=9\) (i.e., 286ms) captures word-level transitions. 
In the frequency domain, these windows analyze frequency relationships at corresponding scales—from narrow-band resonances (i.e., \(k=3\)) to complete spectral envelope structures (i.e., \(k=9\)), with each window capturing progressively broader acoustic patterns in both static features and their dynamics. Empirical analysis indicates that \(k \in \{3, 5, 7, 9\}\) offers the best trade-off between representational diversity and generalization capacity, while avoiding the redundancy or noise sensitivity introduced by larger windows (i.e., \(k=11\)).

\section{Conclusion and Discussion}\label{section5}

We have proposed the first unified framework for ADD under various real-world communication degradations. Our framework explicitly addresses ADD in lossy transmission conditions, including speech codec compression and packet losses. The proposed framework outperforms SOTA baselines, achieving both high detection performance and training efficiency, while substantially improving feature quality and classification separability. Notably, the framework maintains strong robustness across diverse and severe real-world communication degradations without requiring high-fidelity inputs, offering a principled and deployable solution for real-world ADD applications.

\paragraph{Limitations.}
Although we have considered a broad set of speech codecs and PLRs in real-world communication systems, the current simulation does not fully cover other real-world distortions such as jitter, latency, echo, loudspeaker characteristics, mobile noise, and other speech codecs.
Additionally, the framework currently assumes access to 4s audio clips, which may be restrictive in real-time and practical scenarios. 
Simulating more complex real-world communication degradations and using shorter audio clips for fast and high-precision detection are needed. We plan to explore these directions in the future.

\paragraph{Broader impacts.}
The increasing abuse of synthesized speech poses a serious threat to voice authentication systems, digital trust, public safety, and various forms of audio-visual communication. 
Our work contributes a robust framework that enhances the practicality of ADD in real-world telecommunication and security-sensitive domains.
In particular, it offers technical foundations for defending against fraud, voice cloning and mis/disinformation spread via deepfake audio.
However, a potential concern lies in the risk of audio surveillance or data leakage during the detection process. We hope this work encourages the development of ADD towards more practical and deployable solutions in real-world scenarios.

\bibliography{aaai2026}

\begin{thebibliography}{73}
\providecommand{\natexlab}[1]{#1}

\bibitem[{Astrom et~al.(2009)Astrom, Astrom, Spittka, and Vos}]{silk}
Astrom, H.; Astrom, H.; Spittka, J.; and Vos, K. 2009.
\newblock RTP payload format and file storage format for silk speech and audio codec.
\newblock Technical report, Internet Engineering Task Force.

\bibitem[{Besacier et~al.(2003)Besacier, Mayorga, Bonastre, Fredouille, and Meignier}]{besacier2003overview}
Besacier, L.; Mayorga, P.; Bonastre, J.-F.; Fredouille, C.; and Meignier, S. 2003.
\newblock Overview of compression and packet loss effects in speech biometrics.
\newblock \emph{IEE Proceedings-Vision, Image and Signal Processing}, 150(6): 372--376.

\bibitem[{Bessette et~al.(2002)Bessette, Salami, Lefebvre, Jelinek, Rotola-Pukkila, Vainio, Mikkola, and Jarvinen}]{amrwb}
Bessette, B.; Salami, R.; Lefebvre, R.; Jelinek, M.; Rotola-Pukkila, J.; Vainio, J.; Mikkola, H.; and Jarvinen, K. 2002.
\newblock The adaptive multirate wideband speech codec (AMR-WB).
\newblock \emph{IEEE Transactions on Speech and Audio Processing}, 10(8): 620--636.

\bibitem[{Bisogni et~al.(2024)Bisogni, Loia, Nappi, and Pero}]{bisogni2024acoustic}
Bisogni, C.; Loia, V.; Nappi, M.; and Pero, C. 2024.
\newblock Acoustic features analysis for explainable machine learning-based audio spoofing detection.
\newblock \emph{Computer Vision and Image Understanding}, 249: 104145.

\bibitem[{Blue et~al.(2022)Blue, Warren, Abdullah, Gibson, Vargas, O'Dell, Butler, and Traynor}]{ref28}
Blue, L.; Warren, K.; Abdullah, H.; Gibson, C.; Vargas, L.; O'Dell, J.; Butler, K.; and Traynor, P. 2022.
\newblock Who are you (I really wanna know)? detecting audio DeepFakes through vocal tract reconstruction.
\newblock In \emph{31st USENIX Security Symposium (USENIX Security 22)}, 2691--2708.

\bibitem[{Borodin et~al.(2024)Borodin, Kudryavtsev, Korzh, Efimenko, Mkrtchian, Gorodnichev, and Rogov}]{borodin2024aasist3}
Borodin, K.; Kudryavtsev, V.; Korzh, D.; Efimenko, A.; Mkrtchian, G.; Gorodnichev, M.; and Rogov, O.~Y. 2024.
\newblock AASIST3: KAN-enhanced AASIST speech deepfake detection using SSL features and additional regularization for the ASVspoof 2024 Challenge.
\newblock In \emph{Proc. ASVspoof 2024}, 48--55.

\bibitem[{Bottou, Curtis, and Nocedal(2018)}]{earlystop}
Bottou, L.; Curtis, F.~E.; and Nocedal, J. 2018.
\newblock Optimization methods for large-scale machine learning.
\newblock \emph{SIAM review}, 60(2): 223--311.

\bibitem[{Brewster(2021)}]{2020fraudsters}
Brewster, T. 2021.
\newblock Fraudsters cloned company director’s voice in \$35 million heist, police find.
\newblock \url{https://www.forbes.com/sites/thomasbrewster/2021/10/14/huge-bank-fraud-uses-deep-fake-voice-tech-to-steal-millions/}.
\newblock Accessed: 2025-05-12.

\bibitem[{Bruhn et~al.(2015)Bruhn, Pobloth, Schnell, Grill, Gibbs, Miao, J{\"a}rvinen, Laaksonen, Harada, Naka et~al.}]{evs}
Bruhn, S.; Pobloth, H.; Schnell, M.; Grill, B.; Gibbs, J.; Miao, L.; J{\"a}rvinen, K.; Laaksonen, L.; Harada, N.; Naka, N.; et~al. 2015.
\newblock Standardization of the new 3GPP EVS codec.
\newblock In \emph{IEEE International Conference on Acoustics, Speech and Signal Processing (ICASSP)}, 5703--5707.

\bibitem[{Chakravarty and Dua(2024)}]{ref37}
Chakravarty, N.; and Dua, M. 2024.
\newblock A lightweight feature extraction technique for deepfake audio detection.
\newblock \emph{Multimedia Tools and Applications}, 83(26): 67443--67467.

\bibitem[{Chettri(2023)}]{chettri2023clever}
Chettri, B. 2023.
\newblock The clever hans effect in voice spoofing detection.
\newblock In \emph{2022 IEEE Spoken Language Technology Workshop (SLT)}, 577--584. IEEE.

\bibitem[{Cohen et~al.(2022)Cohen, Rimon, Aflalo, and Permuter}]{cohen2022study}
Cohen, A.; Rimon, I.; Aflalo, E.; and Permuter, H.~H. 2022.
\newblock A study on data augmentation in voice anti-spoofing.
\newblock \emph{Speech Communication}, 141: 56--67.

\bibitem[{Coldewey(2024)}]{news3}
Coldewey, D. 2024.
\newblock Six million fine for robocaller who used ai to clone biden’s voice.
\newblock \url{https://techcrunch.com/2024/05/23/6m-fine-for-robocaller-who-used-ai-to-clone-bidens-voice/}.
\newblock Accessed: 2025-05-12.

\bibitem[{Cox(2023)}]{news2}
Cox, J. 2023.
\newblock How i broke into a bank account with an ai-generated voice.
\newblock \url{https://www.vice.com/en/ article/dy7axa/how-i-broke-into-a-bank-account-with-an-ai-generated-voice}.
\newblock Accessed: 2025-05-12.

\bibitem[{Doan et~al.(2023)Doan, Nguyen-Vu, Jung, and Hong}]{ref27}
Doan, T.-P.; Nguyen-Vu, L.; Jung, S.; and Hong, K. 2023.
\newblock BTS-E: Audio Deepfake Detection Using Breathing-Talking-Silence Encoder.
\newblock In \emph{IEEE International Conference on Acoustics, Speech and Signal Processing (ICASSP)}, 1--5.

\bibitem[{ETSI(2024)}]{IVAS}
ETSI. 2024.
\newblock LTE; 5G; Codec for Immersive Voice and Audio Services - Detailed Algorithmic Description incl. RTP payload format and SDP parameter definitions.
\newblock https://www.etsi.org/.

\bibitem[{Frank and Sch\"{o}nherr(2021)}]{wavefake}
Frank, J.; and Sch\"{o}nherr, L. 2021.
\newblock WaveFake: A Data Set to Facilitate Audio Deepfake Detection.
\newblock In \emph{Proceedings of the Neural Information Processing Systems Track on Datasets and Benchmarks}, volume~1.

\bibitem[{Gerken and McMahon(2022)}]{news4}
Gerken, T.; and McMahon, L. 2022.
\newblock Big tech must deal with disinformation or face fines, says eu.
\newblock \url{https://www.bbc.co.uk/news/technology-61817647}.
\newblock Accessed: 2025-05-12.

\bibitem[{Goode(2002)}]{voip}
Goode, B. 2002.
\newblock Voice over internet protocol (voip).
\newblock \emph{Proceedings of the IEEE}, 90(9): 1495--1517.

\bibitem[{Guo et~al.(2024)Guo, Huang, Chen, Zhao, and Wang}]{ref32}
Guo, Y.; Huang, H.; Chen, X.; Zhao, H.; and Wang, Y. 2024.
\newblock Audio Deepfake Detection With Self-Supervised Wavlm And Multi-Fusion Attentive Classifier.
\newblock In \emph{IEEE International Conference on Acoustics, Speech and Signal Processing (ICASSP)}, 12702--12706.

\bibitem[{Hamza et~al.(2022)Hamza, Javed, Iqbal, Kryvinska, Almadhor, Jalil, and Borghol}]{ref35}
Hamza, A.; Javed, A. R.~R.; Iqbal, F.; Kryvinska, N.; Almadhor, A.~S.; Jalil, Z.; and Borghol, R. 2022.
\newblock Deepfake Audio Detection via MFCC Features Using Machine Learning.
\newblock \emph{IEEE Access}, 10: 134018--134028.

\bibitem[{Hu, Shen, and Sun(2018)}]{hu2018squeeze}
Hu, J.; Shen, L.; and Sun, G. 2018.
\newblock Squeeze-and-excitation networks.
\newblock In \emph{Proceedings of the IEEE conference on computer vision and pattern recognition}, 7132--7141.

\bibitem[{Ito and Johnson(2017)}]{ljspeech}
Ito, K.; and Johnson, L. 2017.
\newblock The LJ Speech Dataset.
\newblock https://keithito.com/LJ-Speech-Dataset/.

\bibitem[{Jia et~al.(2016)Jia, De~Brabandere, Tuytelaars, and Gool}]{jia2016dynamic}
Jia, X.; De~Brabandere, B.; Tuytelaars, T.; and Gool, L.~V. 2016.
\newblock Dynamic filter networks.
\newblock \emph{Advances in neural information processing systems}, 29.

\bibitem[{Jung et~al.(2022)Jung, Heo, Tak, Shim, Chung, Lee, Yu, and Evans}]{aasist}
Jung, J.-w.; Heo, H.-S.; Tak, H.; Shim, H.-j.; Chung, J.~S.; Lee, B.-J.; Yu, H.-J.; and Evans, N. 2022.
\newblock Aasist: Audio anti-spoofing using integrated spectro-temporal graph attention networks.
\newblock In \emph{IEEE International Conference on Acoustics, Speech and Signal Processing (ICASSP)}, 6367--6371.

\bibitem[{Kanwal et~al.(2024)Kanwal, Mahum, AlSalman, Sharaf, and Hassan}]{ref29}
Kanwal, T.; Mahum, R.; AlSalman, A.~M.; Sharaf, M.; and Hassan, H. 2024.
\newblock Fake speech detection using VGGish with attention block.
\newblock \emph{EURASIP Journal on Audio, Speech, and Music Processing}, 2024(1): 35.

\bibitem[{Knibbs(2024)}]{news1}
Knibbs, K. 2024.
\newblock Researchers say the deepfake biden robocall was likely made with tools from ai startup elevenlabs.
\newblock \url{https://www.wired.com/story/biden-robocall-deepfake-elevenlabs/}.
\newblock Accessed: 2025-05-12.

\bibitem[{Kong, Kim, and Bae(2020)}]{kong2020hifi}
Kong, J.; Kim, J.; and Bae, J. 2020.
\newblock Hifi-gan: Generative adversarial networks for efficient and high fidelity speech synthesis.
\newblock \emph{Advances in Neural Information Processing Systems}, 33: 17022--17033.

\bibitem[{Kong et~al.(2021)Kong, Ping, Huang, Zhao, and Catanzaro}]{kong2020diffwave}
Kong, Z.; Ping, W.; Huang, J.; Zhao, K.; and Catanzaro, B. 2021.
\newblock DiffWave: A Versatile Diffusion Model for Audio Synthesis.
\newblock In \emph{International Conference on Learning Representations}.

\bibitem[{Kumar et~al.(2019)Kumar, Kumar, De~Boissiere, Gestin, Teoh, Sotelo, De~Brebisson, Bengio, and Courville}]{kumar2019melgan}
Kumar, K.; Kumar, R.; De~Boissiere, T.; Gestin, L.; Teoh, W.~Z.; Sotelo, J.; De~Brebisson, A.; Bengio, Y.; and Courville, A.~C. 2019.
\newblock Melgan: Generative adversarial networks for conditional waveform synthesis.
\newblock \emph{Advances in Neural Information Processing Systems}, 32: 14910 -- 14921.

\bibitem[{Lavrentyeva et~al.(2019)Lavrentyeva, Novoselov, Tseren, Volkova, Gorlanov, and Kozlov}]{lavrentyeva2019stc}
Lavrentyeva, G.; Novoselov, S.; Tseren, A.; Volkova, M.; Gorlanov, A.; and Kozlov, A. 2019.
\newblock STC Antispoofing Systems for the ASVspoof2019 Challenge.
\newblock \emph{Interspeech 2019}.

\bibitem[{Li, Ahmadiadli, and Zhang(2022)}]{ref19}
Li, M.; Ahmadiadli, Y.; and Zhang, X.-P. 2022.
\newblock A comparative study on physical and perceptual features for deepfake audio detection.
\newblock In \emph{Proceedings of the 1st International Workshop on Deepfake Detection for Audio Multimedia}, 35--41.

\bibitem[{Li et~al.(2019)Li, Wang, Hu, and Yang}]{li2019selective}
Li, X.; Wang, W.; Hu, X.; and Yang, J. 2019.
\newblock Selective Kernel Networks.
\newblock In \emph{2019 IEEE/CVF Conference on Computer Vision and Pattern Recognition (CVPR)}. IEEE.

\bibitem[{Lin et~al.(2017)Lin, Doll{\'a}r, Girshick, He, Hariharan, and Belongie}]{lin2017feature}
Lin, T.-Y.; Doll{\'a}r, P.; Girshick, R.; He, K.; Hariharan, B.; and Belongie, S. 2017.
\newblock Feature pyramid networks for object detection.
\newblock In \emph{Proceedings of the IEEE conference on computer vision and pattern recognition}, 2117--2125.

\bibitem[{Liu et~al.(2023)Liu, Wang, Sahidullah, Patino, Delgado, Kinnunen, Todisco, Yamagishi, Evans, Nautsch et~al.}]{asvspoof2021}
Liu, X.; Wang, X.; Sahidullah, M.; Patino, J.; Delgado, H.; Kinnunen, T.; Todisco, M.; Yamagishi, J.; Evans, N.; Nautsch, A.; et~al. 2023.
\newblock Asvspoof 2021: Towards spoofed and deepfake speech detection in the wild.
\newblock \emph{IEEE/ACM Transactions on Audio, Speech, and Language Processing}, 31: 2507--2522.

\bibitem[{Loshchilov and Hutter(2017{\natexlab{a}})}]{adamW}
Loshchilov, I.; and Hutter, F. 2017{\natexlab{a}}.
\newblock Decoupled weight decay regularization.
\newblock \emph{arXiv preprint arXiv:1711.05101}.

\bibitem[{Loshchilov and Hutter(2017{\natexlab{b}})}]{cos}
Loshchilov, I.; and Hutter, F. 2017{\natexlab{b}}.
\newblock {SGDR}: Stochastic Gradient Descent with Warm Restarts.
\newblock In \emph{International Conference on Learning Representations}.

\bibitem[{M, Rajput, and M(2024)}]{ref2}
M, S.; Rajput, A.; and M, S. 2024.
\newblock Classification of Deep Fake Audio Using MFCC Technique.
\newblock In \emph{IEEE International Conference on Information Technology, Electronics and Intelligent Communication Systems (ICITEICS)}, 1--6.

\bibitem[{{M-AILABS}(2019)}]{mailabs}
{M-AILABS}. 2019.
\newblock The M-AILABS Speech Dataset.
\newblock https://github.com/imdatceleste/m-ailabs-dataset.

\bibitem[{Martín-Doñas and Álvarez(2022)}]{ref22}
Martín-Doñas, J.~M.; and Álvarez, A. 2022.
\newblock The Vicomtech Audio Deepfake Detection System Based on Wav2vec2 for the 2022 ADD Challenge.
\newblock In \emph{IEEE International Conference on Acoustics, Speech and Signal Processing (ICASSP)}, 9241--9245.

\bibitem[{Molisch(2012)}]{molisch2012wireless}
Molisch, A.~F. 2012.
\newblock \emph{Wireless communications}, volume~34.
\newblock John Wiley \& Sons.

\bibitem[{M{\"u}ller et~al.(2024)M{\"u}ller, Kawa, Choong, Casanova, G{\"o}lge, M{\"u}ller, Syga, Sperl, and B{\"o}ttinger}]{mlaad}
M{\"u}ller, N.~M.; Kawa, P.; Choong, W.~H.; Casanova, E.; G{\"o}lge, E.; M{\"u}ller, T.; Syga, P.; Sperl, P.; and B{\"o}ttinger, K. 2024.
\newblock Mlaad: The multi-language audio anti-spoofing dataset.
\newblock In \emph{2024 International Joint Conference on Neural Networks (IJCNN)}, 1--7.

\bibitem[{Reimao and Tzerpos(2019)}]{FoR}
Reimao, R.; and Tzerpos, V. 2019.
\newblock For: A dataset for synthetic speech detection.
\newblock In \emph{2019 International Conference on Speech Technology and Human-Computer Dialogue (SpeD)}, 1--10.

\bibitem[{Ren et~al.(2021)Ren, Hu, Tan, Qin, Zhao, Zhao, and Liu}]{ren2020fastspeech}
Ren, Y.; Hu, C.; Tan, X.; Qin, T.; Zhao, S.; Zhao, Z.; and Liu, T.-Y. 2021.
\newblock FastSpeech 2: Fast and High-Quality End-to-End Text to Speech.
\newblock In \emph{International Conference on Learning Representations}.

\bibitem[{Sahidullah et~al.(2025)Sahidullah, Shim, Hautam{\"a}ki, and Kinnunen}]{sahidullah2025shortcut}
Sahidullah, M.; Shim, H.-j.; Hautam{\"a}ki, R.~G.; and Kinnunen, T.~H. 2025.
\newblock Shortcut Learning in Binary Classifier Black Boxes: Applications to Voice Anti-Spoofing and Biometrics.
\newblock \emph{IEEE Journal of Selected Topics in Signal Processing}.

\bibitem[{Sesia, Toufik, and Baker(2011)}]{volte}
Sesia, S.; Toufik, I.; and Baker, M. 2011.
\newblock \emph{Lte-the umts long term evolution: from theory to practice}.
\newblock Wiley.

\bibitem[{Shen et~al.(2018)Shen, Pang, Weiss, Schuster, Jaitly, Yang, Chen, Zhang, Wang, Skerrv-Ryan et~al.}]{shen2018natural}
Shen, J.; Pang, R.; Weiss, R.~J.; Schuster, M.; Jaitly, N.; Yang, Z.; Chen, Z.; Zhang, Y.; Wang, Y.; Skerrv-Ryan, R.; et~al. 2018.
\newblock Natural tts synthesis by conditioning wavenet on mel spectrogram predictions.
\newblock In \emph{IEEE International Conference on Acoustics, Speech and Signal Processing (ICASSP)}, 4779--4783.

\bibitem[{Shi, Shi, and Dogan(2024)}]{shi2024}
Shi, H.; Shi, X.; and Dogan, S. 2024.
\newblock Speech inpainting based on multi-layer long short-term memory networks.
\newblock \emph{Future Internet}, 16(2): 63.

\bibitem[{Shi et~al.(2025)Shi, Shi, Dogan, Alzubi, Huang, and Zhang}]{haohan}
Shi, H.; Shi, X.; Dogan, S.; Alzubi, S.; Huang, T.; and Zhang, Y. 2025.
\newblock Benchmarking Audio Deepfake Detection Robustness in Real-world Communication Scenarios.
\newblock \emph{arXiv preprint arXiv:2504.12423}.
\newblock Accepted by EUSIPCO 2025.

\bibitem[{Shih, Yeh, and Chen(2024)}]{shih2024does}
Shih, T.-H.; Yeh, C.-Y.; and Chen, M.-S. 2024.
\newblock Does Audio Deepfake Detection Rely on Artifacts?
\newblock In \emph{ICASSP 2024-2024 IEEE International Conference on Acoustics, Speech and Signal Processing (ICASSP)}, 12446--12450. IEEE.

\bibitem[{Shim et~al.(2023)Shim, Gonzalez~Hautam{\"a}ki, Sahidullah, and Kinnunen}]{shim2023construct}
Shim, H.-j.; Gonzalez~Hautam{\"a}ki, R.; Sahidullah, M.; and Kinnunen, T. 2023.
\newblock How to Construct Perfect and Worse-than-Coin-Flip Spoofing Countermeasures: A Word of Warning on Shortcut Learning.
\newblock In \emph{Proc. Interspeech 2023}, 785--789.

\bibitem[{Tak et~al.(2021{\natexlab{a}})Tak, Jung, Patino, Kamble, Todisco, and Evans}]{rawgat}
Tak, H.; Jung, J.-W.; Patino, J.; Kamble, M.; Todisco, M.; and Evans, N. 2021{\natexlab{a}}.
\newblock End-to-End Spectro-Temporal Graph Attention Networks for Speaker Verification Anti-Spoofing and Speech Deepfake Detection.
\newblock In \emph{ASVSPOOF 2021, Automatic Speaker Verification and Spoofing Countermeasures Challenge}, 1--8. ISCA.

\bibitem[{Tak et~al.(2021{\natexlab{b}})Tak, Patino, Todisco, Nautsch, Evans, and Larcher}]{rawnet2}
Tak, H.; Patino, J.; Todisco, M.; Nautsch, A.; Evans, N.; and Larcher, A. 2021{\natexlab{b}}.
\newblock End-to-end anti-spoofing with rawnet2.
\newblock In \emph{IEEE International Conference on Acoustics, Speech and Signal Processing (ICASSP)}, 6369--6373.

\bibitem[{Tak et~al.(2022)Tak, Todisco, Wang, Jung, Yamagishi, and Evans}]{ref39}
Tak, H.; Todisco, M.; Wang, X.; Jung, J.-w.; Yamagishi, J.; and Evans, N. 2022.
\newblock Automatic speaker verification spoofing and deepfake detection using wav2vec 2.0 and data augmentation.
\newblock \emph{arXiv preprint arXiv:2202.12233}.

\bibitem[{Todisco, Delgado, and Evans(2017)}]{todisco2017constant}
Todisco, M.; Delgado, H.; and Evans, N. 2017.
\newblock Constant Q cepstral coefficients: A spoofing countermeasure for automatic speaker verification.
\newblock \emph{Computer Speech \& Language}, 45: 516--535.

\bibitem[{Todisco et~al.(2019)Todisco, Wang, Vestman, Sahidullah, Delgado, Nautsch, Yamagishi, Evans, Kinnunen, and Lee}]{asvspoof2019}
Todisco, M.; Wang, X.; Vestman, V.; Sahidullah, M.; Delgado, H.; Nautsch, A.; Yamagishi, J.; Evans, N.; Kinnunen, T.; and Lee, K.~A. 2019.
\newblock ASVspoof 2019: Future horizons in spoofed and fake audio detection.
\newblock \emph{arXiv preprint arXiv:1904.05441}.

\bibitem[{Valenti et~al.(2017)Valenti, Squartini, Diment, Parascandolo, and Virtanen}]{valenti2017convolutional}
Valenti, M.; Squartini, S.; Diment, A.; Parascandolo, G.; and Virtanen, T. 2017.
\newblock A convolutional neural network approach for acoustic scene classification.
\newblock In \emph{2017 International Joint Conference on Neural Networks (IJCNN)}, 1547--1554. IEEE.

\bibitem[{Valin(2016)}]{speex}
Valin, J.-M. 2016.
\newblock Speex: A free codec for free speech.
\newblock \emph{arXiv preprint arXiv:1602.08668}.

\bibitem[{Valin et~al.(2016)Valin, Maxwell, Terriberry, and Vos}]{opus}
Valin, J.-M.; Maxwell, G.; Terriberry, T.~B.; and Vos, K. 2016.
\newblock High-quality, low-delay music coding in the opus codec.
\newblock \emph{arXiv preprint arXiv:1602.04845}.

\bibitem[{Van Den~Oord et~al.(2016)Van Den~Oord, Dieleman, Zen, Simonyan, Vinyals, Graves, Kalchbrenner, Senior, Kavukcuoglu et~al.}]{van2016wavenet}
Van Den~Oord, A.; Dieleman, S.; Zen, H.; Simonyan, K.; Vinyals, O.; Graves, A.; Kalchbrenner, N.; Senior, A.; Kavukcuoglu, K.; et~al. 2016.
\newblock Wavenet: A generative model for raw audio.
\newblock \emph{arXiv preprint arXiv:1609.03499}, 12.

\bibitem[{Van~der Maaten and Hinton(2008)}]{tsne}
Van~der Maaten, L.; and Hinton, G. 2008.
\newblock Visualizing data using t-SNE.
\newblock \emph{Journal of Machine Learning Research}, 9(11): 2579--2605.

\bibitem[{Wang et~al.(2024)Wang, Delgado, Tak, Jung, Shim, Todisco, Kukanov, Liu, Sahidullah, Kinnunen et~al.}]{wang2024asvspoof}
Wang, X.; Delgado, H.; Tak, H.; Jung, J.-w.; Shim, H.-j.; Todisco, M.; Kukanov, I.; Liu, X.; Sahidullah, M.; Kinnunen, T.; et~al. 2024.
\newblock ASVspoof 5: Crowdsourced speech data, deepfakes, and adversarial attacks at scale.
\newblock \emph{arXiv preprint arXiv:2408.08739}.

\bibitem[{Wang et~al.(2018)Wang, Girshick, Gupta, and He}]{wang2018non}
Wang, X.; Girshick, R.; Gupta, A.; and He, K. 2018.
\newblock Non-local neural networks.
\newblock In \emph{Proceedings of the IEEE conference on computer vision and pattern recognition}, 7794--7803.

\bibitem[{Wang and Yamagishi(2021)}]{ref23}
Wang, X.; and Yamagishi, J. 2021.
\newblock Investigating self-supervised front ends for speech spoofing countermeasures.
\newblock \emph{arXiv preprint arXiv:2111.07725}.

\bibitem[{Wani et~al.(2024{\natexlab{a}})Wani, Qadri, Comminiello, and Amerini}]{ref24}
Wani, T.~M.; Qadri, S. A.~A.; Comminiello, D.; and Amerini, I. 2024{\natexlab{a}}.
\newblock Detecting audio deepfakes: Integrating CNN and BiLSTM with multi-feature concatenation.
\newblock In \emph{Proceedings of the 2024 ACM Workshop on Information Hiding and Multimedia Security}, 271--276.

\bibitem[{Wani et~al.(2024{\natexlab{b}})Wani, Qadri, Comminiello, and Amerini}]{ref7}
Wani, T.~M.; Qadri, S. A.~A.; Comminiello, D.; and Amerini, I. 2024{\natexlab{b}}.
\newblock Detecting audio deepfakes: Integrating CNN and BiLSTM with multi-feature concatenation.
\newblock In \emph{Proceedings of the 2024 ACM Workshop on Information Hiding and Multimedia Security}, 271--276.

\bibitem[{Woo et~al.(2018)Woo, Park, Lee, and Kweon}]{woo2018cbam}
Woo, S.; Park, J.; Lee, J.-Y.; and Kweon, I.-S. 2018.
\newblock CBAM: Convolutional Block Attention Module.
\newblock In \emph{European Conference on Computer Vision}, 3--19. European Conference on Computer Vision.

\bibitem[{Yadav and Rai(2020)}]{yadav2020frequency}
Yadav, S.; and Rai, A. 2020.
\newblock Frequency and temporal convolutional attention for text-independent speaker recognition.
\newblock In \emph{ICASSP 2020-2020 IEEE international conference on acoustics, speech and signal processing (ICASSP)}, 6794--6798. IEEE.

\bibitem[{Yamamoto, Song, and Kim(2020)}]{yamamoto2020parallel}
Yamamoto, R.; Song, E.; and Kim, J.-M. 2020.
\newblock Parallel WaveGAN: A fast waveform generation model based on generative adversarial networks with multi-resolution spectrogram.
\newblock In \emph{IEEE International Conference on Acoustics, Speech and Signal Processing (ICASSP)}, 6199--6203.

\bibitem[{Yi et~al.(2022)Yi, Fu, Tao, Nie, Ma, Wang, Wang, Tian, Bai, Fan et~al.}]{add2022}
Yi, J.; Fu, R.; Tao, J.; Nie, S.; Ma, H.; Wang, C.; Wang, T.; Tian, Z.; Bai, Y.; Fan, C.; et~al. 2022.
\newblock Add 2022: the first audio deep synthesis detection challenge.
\newblock In \emph{IEEE International Conference on Acoustics, Speech and Signal Processing (ICASSP)}, 9216--9220.

\bibitem[{Yu et~al.(2024)Yu, Chen, Leng, Chen, and Yi}]{ref49}
Yu, N.; Chen, L.; Leng, T.; Chen, Z.; and Yi, X. 2024.
\newblock An explainable deepfake of speech detection method with spectrograms and waveforms.
\newblock \emph{Journal of Information Security and Applications}, 81: 103720.

\bibitem[{Zhang, Wen, and Hu(2024)}]{ref15}
Zhang, Q.; Wen, S.; and Hu, T. 2024.
\newblock Audio deepfake detection with self-supervised xls-r and sls classifier.
\newblock In \emph{Proceedings of the 32nd ACM International Conference on Multimedia}, 6765--6773.

\bibitem[{Zhu et~al.(2024)Zhu, Koppisetti, Tran, and Bharaj}]{zhu2024slim}
Zhu, Y.; Koppisetti, S.; Tran, T.; and Bharaj, G. 2024.
\newblock Slim: Style-linguistics mismatch model for generalized audio deepfake detection.
\newblock \emph{Advances in Neural Information Processing Systems}, 37: 67901--67928.

\end{thebibliography}



\clearpage
\newpage

\appendix
\section*{Appendix}

\section{Supplemental details of Figure \ref{fig1} and feature dispersion analysis}\label{fig1details}



The audio samples used in Figure \ref{fig1} are randomly selected from the ADD-C test dataset. Details of the ADD-C test dataset are provided in Appendix \ref{addcdetails}. For the clean condition, 1,000 real and 1,000 fake utterances are sampled from \(C_0\), which contains only high-fidelity audio signals unaffected by speech codec compression or packet losses. 
For the communication degraded condition, we randomly sample 200 real and 200 fake utterances from each of the five conditions (\(C_1\) to \(C_5\)) and aggregate them into a balanced subset comprising 1,000 real and 1,000 fake utterances. This subset spans all six speech codecs and five packet loss levels, offering a comprehensive representation of real-world communication degradations.

Figure \ref{fig1} shows the distributions of real and fake audio samples without (top row) and with (bottom row) real-world communication degradations, including speech codec compression and packet losses.
The visualization difference highlights how real-world communication degradations impact the original feature structure and increase intra-class dispersion and class boundaries, making detection more challenging.
Another notable observation from Figure \ref{fig1} is the expansion of the horizontal and vertical axis ranges with the communication degraded effects. This spread reflects weakened clustering structures and increased feature dispersion due to real-world communication degradations, thereby making the ADD task significantly harder compared to clean input.
Additionally, the marginal distributions further validate the observation, showing higher density peaks in clean conditions where samples form distinct clusters, whereas communication degradation leads to flatter distributions with lower peak values, which is the quantitative evidence of feature dispersion and class boundary deterioration.

\begin{figure}[h]
  \centering
  \includegraphics[width=0.45\textwidth]{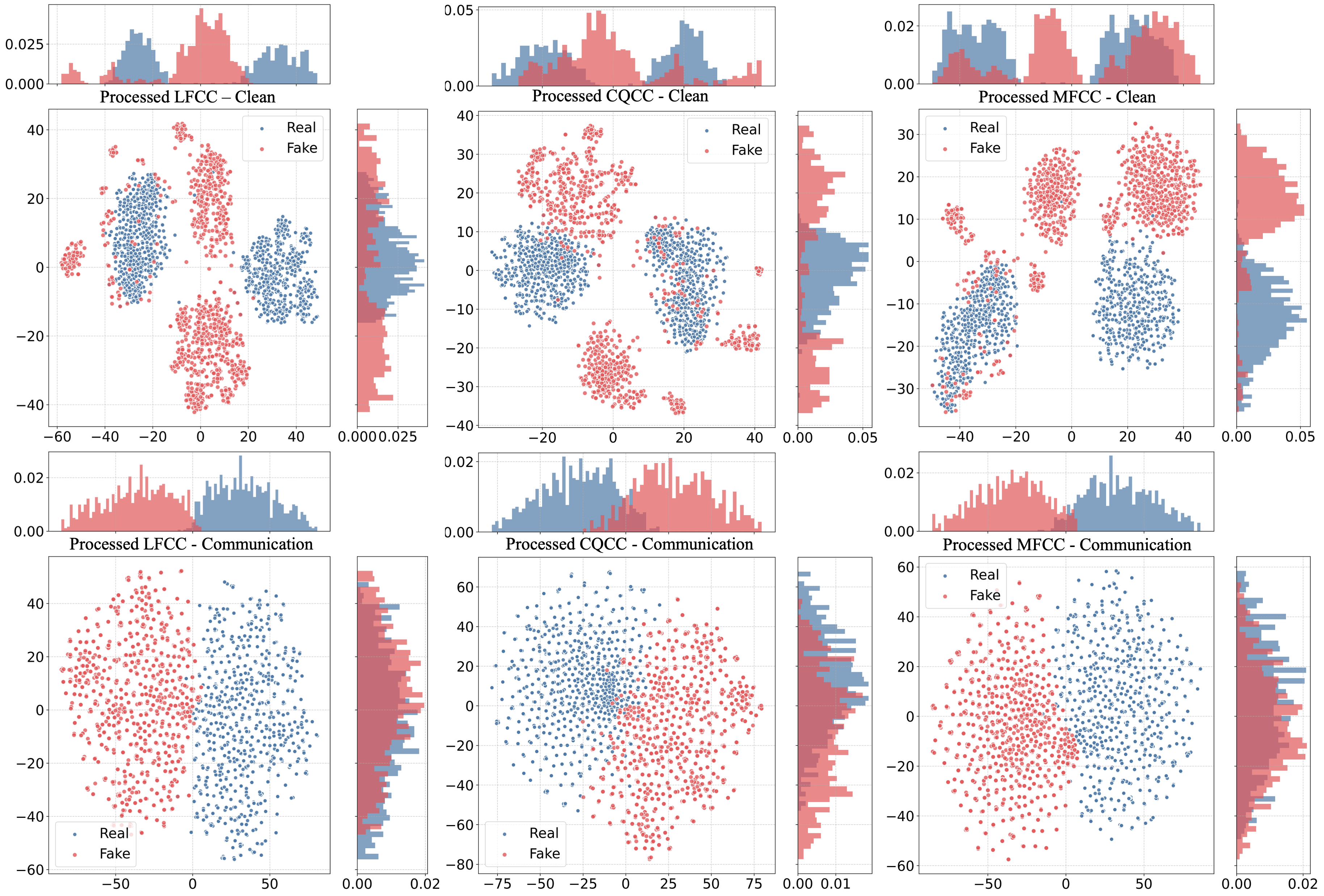}
  \caption{t-SNE visualizations of real and fake audio samples corresponding to Figure \ref{fig1} after being processed by the proposed framework.}
  \label{fig1processed}
\end{figure}

To further illustrate the effectiveness of our framework, we provide the t-SNE visualizations for the processed TF features embeddings under Clean and Communication corresponding to Figure \ref{fig1}, as shown in Figure \ref{fig1processed}, all features are extracted from the proposed framework before the Classifier. The processed feature embeddings exhibit well-formed clusters and clearly separated decision boundaries under both conditions, confirming the framework’s robustness and effectiveness in improving and enhancing discriminative structures from both clean and real-world communication degraded inputs.


\section{Supplemental details of experiments setup}\label{appenB}

\subsection{Construction of \(\mathcal{D}\) and \(\mathcal{D}_{com}\)}\label{dataset}


The details of the datasets used to construct \(\mathcal{D}_{com}\) are shown in Table \ref{tabdataset}. The LJSpeech and M-AILABS datasets contain only real utterances, whereas WaveFake and MLAAD-EN contain only fake ones. Notably, the Wavefake dataset is generated based on the LJSpeech dataset, and MLAAD-EN is also generated based on M-AILABS. The FoR and ASVLA datasets include both real and fake utterances. All audio signals are converted to a single-channel 16-bit Pulse-Code Modulation format with a sampling rate of 16kHz.

\begin{table*}[h]
\centering
\setlength{\tabcolsep}{5mm}
\renewcommand{\arraystretch}{1.1}
\scalebox{0.86}{\begin{tabular}{ccccc}
\toprule
Dataset& Real & Fake & Language & Algorithms  \\ \midrule
FoR \cite{FoR}&34605 &34695 & English &7\\
Wavefake \cite{wavefake}& - & 91700 &English &7 \\
LJSpeech \cite{ljspeech}& 13100 & - &English &-\\
MLAAD-EN \cite{mlaad} & - &  5000& English &5\\
M-AILABS \cite{mailabs} & 69853 &  -& English &-\\
ASVLA \cite{asvspoof2021}& 12483 & 108978 & English &17\\ \midrule
Total & 130041 & 240373 & - & 36 \\
\bottomrule
\end{tabular}}
\caption{Details of the selected six publicly available speech datasets.\label{tabdataset}}
\end{table*}

\(\mathcal{D}\) was constructed by aggregating all real and fake utterances from the six datasets. To construct \(\mathcal{D}_{com}\), we adopted the data augmentation strategy proposed in \cite{haohan}. Specifically, \(\mathcal{D}\) was randomly and proportionally divided into six subsets, each of which was subsequently processed using one of the six speech codecs listed in Table \ref{codecdetail} to simulate codec compression. These subsets were then merged to form a single dataset, which was further augmented using a packet loss simulator to simulate real-world lossy transmission degradation. This resulted in five additional augmented datasets with PLR of 0\%, 1\%, 5\%, 10\%, and 20\%, respectively. Each augmented dataset corresponds to one PLR and contains the codec-introduced compression of six speech codecs. Finally, these augmented datasets were merged to form the final training dataset \(\mathcal{D}_{com}\). The size of \(\mathcal{D}_{com}\) is five times that of \(\mathcal{D}\), greatly enriching the training corpus and covering 30 types of real-world communication degradations. Note that \(\mathcal{D}_{com}\) contains no high-fidelity (Clean) audio signal in its training data, as all utterances are degraded by real-world communication effects.

Table \ref{codecdetail} lists the details of the selected speech codecs and their corresponding settings, including sample rate and bitrate. It is worth noting that the selected AMR-WB speech codec supports a maximum bitrate of 23.85kbps, while other codecs use the closest bitrate of 24.40kbps.

\begin{table*}[h]
\centering
\setlength{\tabcolsep}{1.4mm}
\renewcommand{\arraystretch}{1.05}
\scalebox{0.8}{\begin{tabular}{ccccc}
\toprule
Codec & Support Sample Rate(kHz) & Selected Sample Rate(kHz) &Support Bitrate(kbps) &Selected Bitrate(kbps)  \\ \midrule
AMR-WB \cite{amrwb}& 16 & 16 & 6.60-23.85& 23.85 \\
EVS \cite{evs}& 8,16,32,48 & 16 & 5.90-128 & 24.40 \\
IVAS \cite{IVAS}& 8,16,32,48 & 16 & 13.20-512& 24.40\\
OPUS \cite{opus}& 8-48 & 16 & 6-510& 24.40\\ 
Speex(WB) \cite{speex}& 8,16,32& 16 & 2-44& 24.40\\
SILK \cite{silk} & 8-24& 16 & 6-40& 24.40\\
\bottomrule
\end{tabular}}
\caption{Details of the selected speech codec and settings.\label{codecdetail}}
\end{table*}

\subsection{ADD-C test dataset}\label{addcdetails}


The construction of the ADD-C test dataset follows the protocol outlined in \cite{haohan}, with details listed in Table \ref{addcdetailstab}.
ADD-C includes six distinct conditions \(C_0\)-\(C_5\). \(C_0\) represents clean audio data and consists of 2000 real and 2000 fake utterances. Specifically, 500 fake utterances are randomly selected from each of the WaveFake and MLAAD-EN datasets, while 500 real utterances are randomly selected from each of the LJSpeech and M-AILABS datasets. An additional 500 real and 500 fake utterances are randomly selected from each of the FoR and ASVLA datasets. This selection strategy ensures that both real and fake utterances originate from four different source datasets, thereby enhancing data diversity and ensuring the robustness of evaluation outcomes.
\(C_1\) to \(C_5\) are derived from \(C_0\) by applying various simulated real-world communication degradations. 
Specifically, for \(C_1\), all clean and fake utterances from \(C_0\) are processed using each of the six speech codecs under PLR of 0\% to introduce both codec compression and packet losses. This results in a total of 12,000 real and 12,000 fake utterances for \(C_1\).
The same process is repeated for \(C_2\) to \(C_5\) with PLR of 1\%, 5\%, 10\%, and 20\%, respectively.
Therefore, each condition from  \(C_1\) to \(C_5\) yields 12,000 real and 12,000 fake utterances, covering six codec-introduced compression with a specific PLR, as presented in Table \ref{codecdetail}.

In summary, the ADD-C dataset spans six conditions, ranging from clean (\(C_0\)) to increasingly severe degradation (\(C_1\)-\(C_5\)), encompasses 124,000 utterances and 30 types of real-world communication degradations. This extensive and diverse test dataset provides a comprehensive measure for assessing the robustness and effectiveness of ADD methods under clean conditions and real-world communication degradations.

\begin{table*}[h]
\centering
\setlength{\tabcolsep}{6mm}
\renewcommand{\arraystretch}{1.1}
\scalebox{0.8}{
\begin{tabular}{lcccccc}
\toprule
\textbf{Condition} & \(C_{0}\) & \(C_{1}\) & \(C_{2}\) & \(C_{3}\) & \(C_{4}\) & \(C_{5}\) \\ \midrule
PLR(\%) & - & 0 & 1 & 5 & 10 & 20 \\ \midrule
\textbf{Real utterances} & \textbf{2000} & \textbf{12000} & \textbf{12000} & \textbf{12000} & \textbf{12000} & \textbf{12000}\\
\multicolumn{1}{l}{\(\hookrightarrow\)Clean}& 2000  & - & - & - & - & -\\
\multicolumn{1}{l}{\(\hookrightarrow\) AMR-WB} & - & 2000 & 2000 & 2000 & 2000 & 2000 \\
\multicolumn{1}{l}{\(\hookrightarrow\) EVS} & - & 2000 & 2000 & 2000 & 2000 & 2000 \\
\multicolumn{1}{l}{\(\hookrightarrow\) IVAS} & - & 2000 & 2000 & 2000 & 2000 & 2000 \\
\multicolumn{1}{l}{\(\hookrightarrow\) OPUS} & - & 2000 & 2000 & 2000 & 2000 & 2000\\
\multicolumn{1}{l}{\(\hookrightarrow\) Speex(WB)} & - & 2000 & 2000 & 2000 & 2000 & 2000 \\ 
\multicolumn{1}{l}{\(\hookrightarrow\) SILK} & - & 2000 & 2000 & 2000 & 2000 & 2000\\\midrule
\textbf{Fake utterances} & \textbf{2000} & \textbf{12000} & \textbf{12000} & \textbf{12000} & \textbf{12000} & \textbf{12000} \\
\multicolumn{1}{l}{\(\hookrightarrow\) Clean}& 2000  & - & - & - & - & - \\
\multicolumn{1}{l}{\(\hookrightarrow\) AMR-WB}& -& 2000 & 2000 & 2000 & 2000 & 2000\\
\multicolumn{1}{l}{\(\hookrightarrow\) EVS} &- & 2000 & 2000 & 2000 & 2000 & 2000\\
\multicolumn{1}{l}{\(\hookrightarrow\) IVAS} & - & 2000 & 2000 & 2000 & 2000 & 2000\\
\multicolumn{1}{l}{\(\hookrightarrow\) OPUS} & -& 2000 & 2000 & 2000 & 2000 & 2000 \\
\multicolumn{1}{l}{\(\hookrightarrow\) Speex(WB)} & -& 2000 & 2000 & 2000 & 2000 & 2000\\ 
\multicolumn{1}{l}{\(\hookrightarrow\) SILK} & - & 2000 & 2000 & 2000 & 2000 & 2000\\\midrule
\textbf{Total utterances} & \textbf{4000} & \textbf{24000} & \textbf{24000} & \textbf{24000} & \textbf{24000} & \textbf{24000} \\
\bottomrule
\end{tabular}}
\caption{Detailed composition of ADD-C test dataset.\label{addcdetailstab}}
\end{table*}

\section{Supplemental details for feature combination study}\label{multifeature}

The effects of combining different TF representations as input features are examined in Table~\ref{featurestudy}. Randomly pairing any two of the three TF features (LFCC, CQCC, MFCC) yields severe degradation on detection performance, while using all three features together actually performs worse than some two-feature combinations.

We attribute this to increased input redundancy and misalignment across different cepstral domains, which may disrupt the attention mechanism or cause feature conflicts during training. These findings indicate that simply concatenating different TF features does not ensure better results, and a more principled fusion strategy is required to achieve further improvements. We plan to address this in future work.

\begin{table}[htb]
\centering
\setlength{\tabcolsep}{0.5mm}
\scalebox{0.9}{
\begin{tabular}{cccccccccccc}
\toprule
\multicolumn{3}{c}{Feature}  & \multicolumn{7}{c}{EER(\%) \(\downarrow\)} \\ \cmidrule(r){1-3} \cmidrule(r){4-10} LFCC & CQCC & MFCC & \(C_{0}\)  & \(C_{1}\) & \(C_{2}\) & \(C_{3}\) & \(C_{4}\) & \(C_{5}\) & Avg.\\
\midrule
\(\bullet\) & \(\bullet\) &\(\times\) & 46.75 & 45.49 & 45.50 & 45.47 & 45.57 & 45.83 & 45.77\\
\(\bullet\) & \(\times\) &\(\bullet\) & 44.75 & 47.44 & 47.52 & 47.89 & 47.97 & 48.54 & 47.35\\
\(\times\) & \(\bullet\) &\(\bullet\) & 38.49 & 41.04 & 41.14 & 40.89 & 41.72 & 42.04 & 40.89\\
\(\bullet\) & \(\bullet\) &\(\bullet\) & 46.14 & 44.31 & 44.21 & 44.07 & 43.94 & 43.64 & 44.39\\
\bottomrule
\end{tabular}}
\caption{Performance Comparison of Different TF Feature Combinations.}
\label{featurestudy}
\end{table}

\section{Extended codec-specific analysis}\label{codectsneAnalysis}







\begin{figure*}[h]
  \centering
  \includegraphics[width=0.8\textwidth]{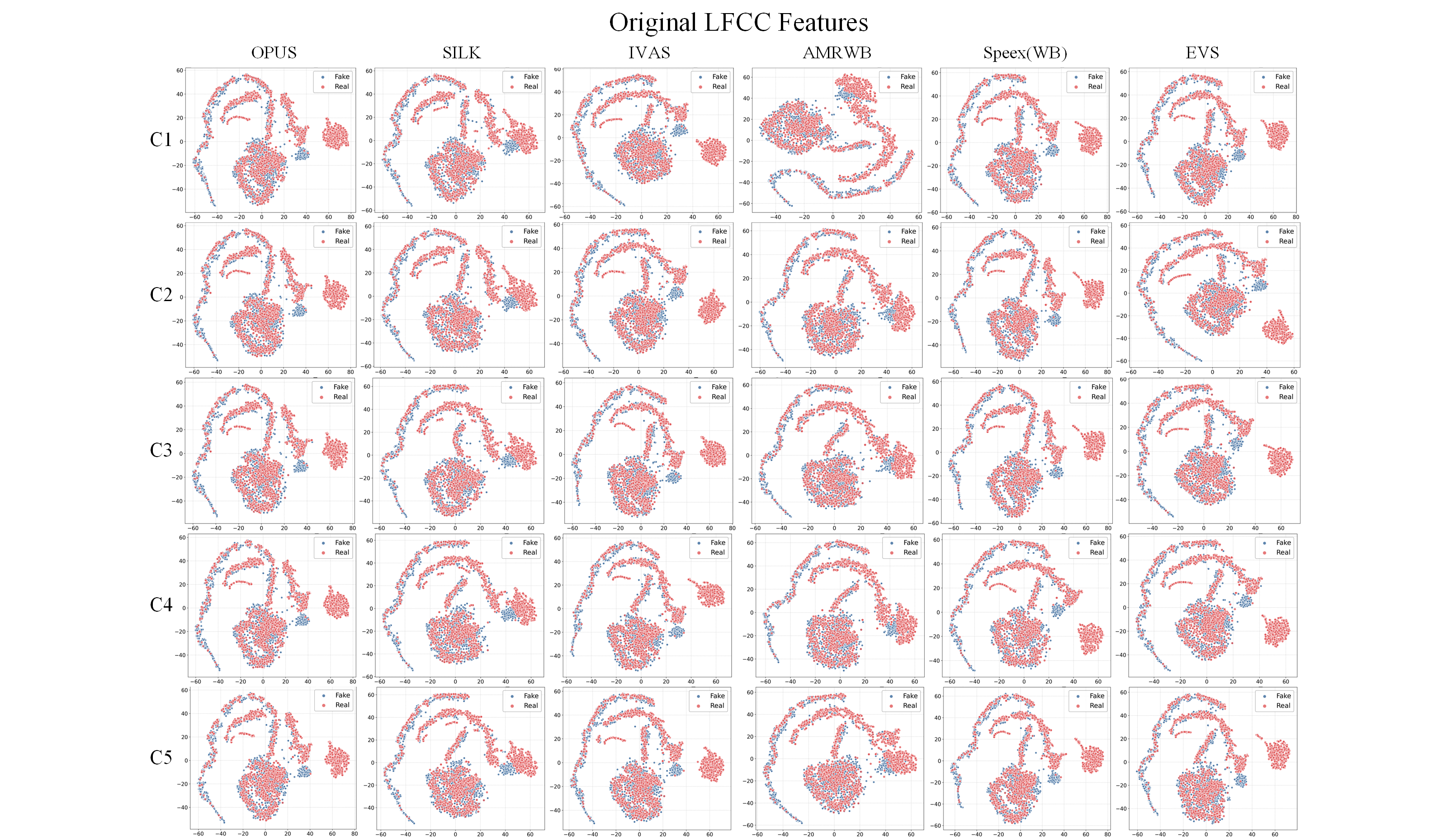}
  \caption{t-SNE visualizations of real and fake audio samples using the original LFCC features, under six speech codecs (columns) and across \(C_1\) to \(C_5\) (rows).}
  \label{appenCodeclfccori}
\end{figure*}
\begin{figure*}[h]
  \centering
  \includegraphics[width=0.8\textwidth]{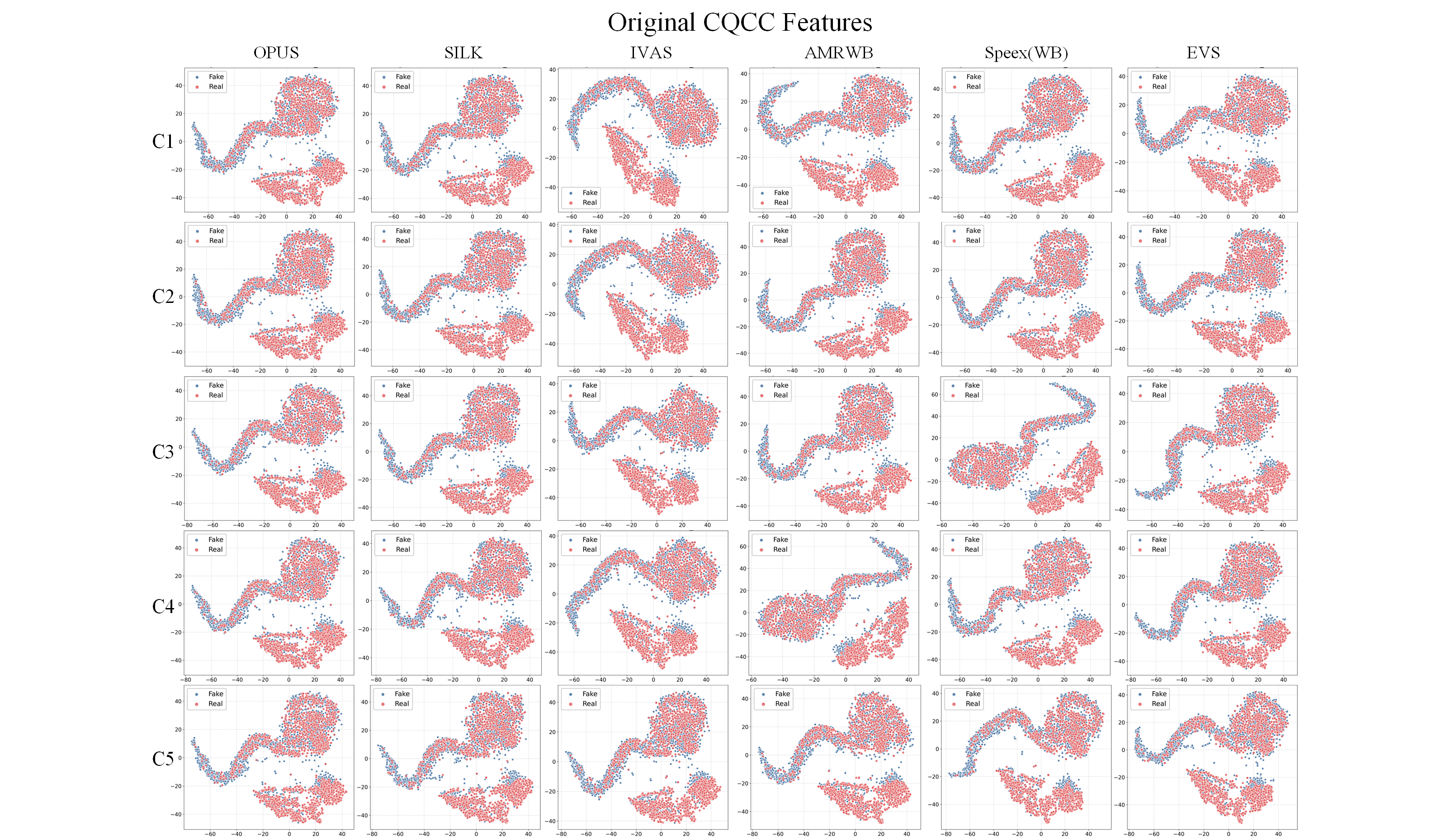}
  \caption{t-SNE visualizations of real and fake audio samples using the original CQCC features, under six speech codecs (columns) and across \(C_1\) to \(C_5\) (rows).}
  \label{appenCodeccqccori}
\end{figure*}
\begin{figure*}[h]
  \centering
  \includegraphics[width=0.8\textwidth]{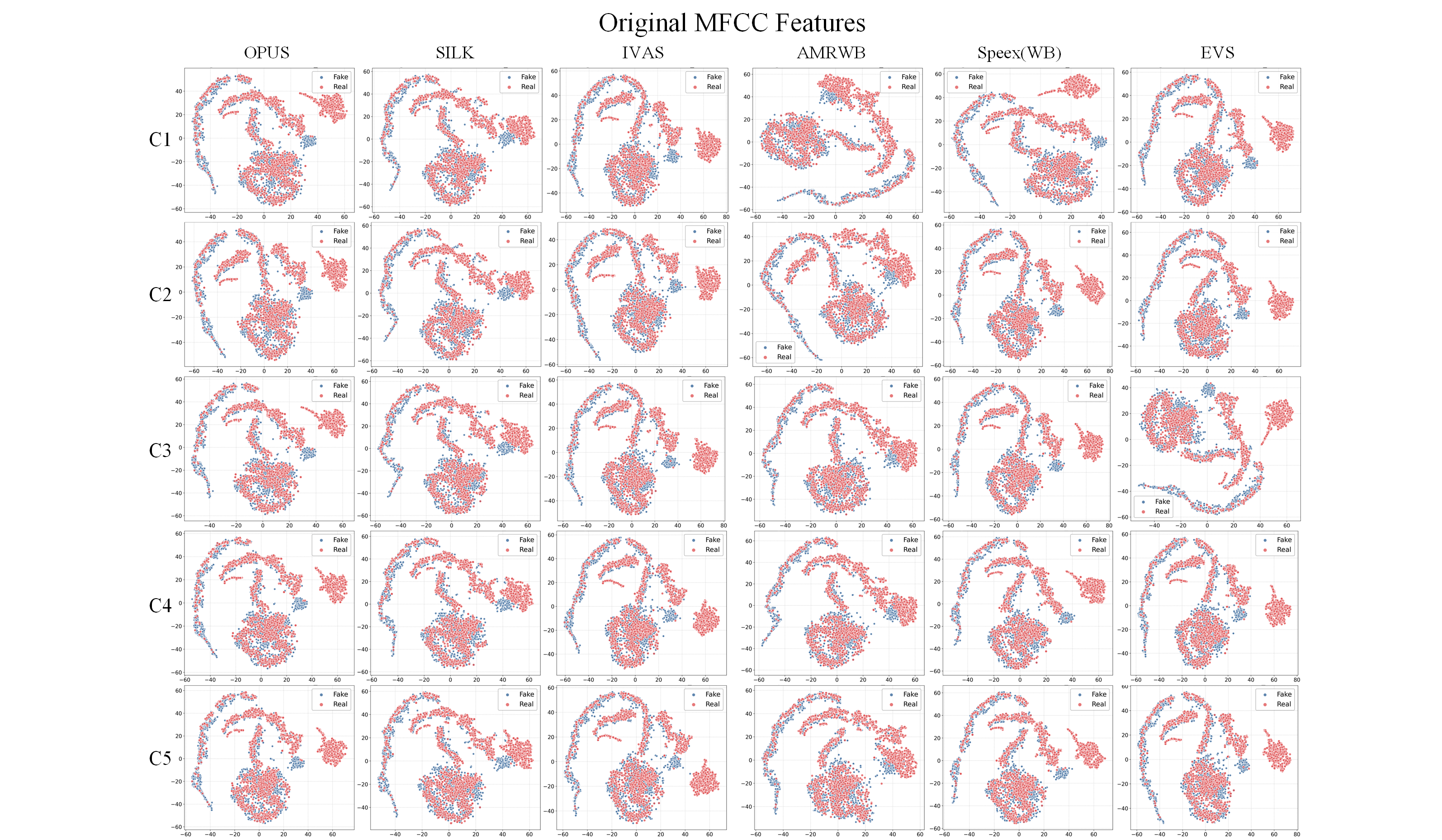}
  \caption{t-SNE visualizations of real and fake audio samples using the original MFCC features, under six speech codecs (columns) and across \(C_1\) to \(C_5\) (rows).}
  \label{appenCodecmfccori}
\end{figure*}

\textbf{This section is to extend and support the codec-specific analysis presented in Section \ref{4.2.2}. }We analyze the underlying architectures of different speech codecs in detail and how these architectures affect the preservation or distortion of differentiated TF features in the case of speech codec compression and packet losses.
t-SNE was employed to show a comprehensive visualization of real and fake audio samples on the original LFCC, CQCC and MFCC features, respectively. The different feature distributions across different speech codecs and conditions are shown in Figure \ref{appenCodeclfccori}, \ref{appenCodeccqccori}, and \ref{appenCodecmfccori}.

OPUS combines Linear Predictive Coding (LPC) and Constrained Energy Lapped Transform (CELT). This hybrid architecture enables dynamic switching or fusion between time and frequency-domain coding according to signal characteristics. 
As shown in the first columns of Figure \ref{appenCodeclfccori}, \ref{appenCodeccqccori}, and \ref{appenCodecmfccori}, the t-SNE distributions exhibit negligible deformation across \(C_1\) to \(C_5\). This indicates that the LPC and CELT can effectively preserve subtle high-resolution TF features and harmonic structures, which are essential for distinguishing real and fake speech under various communication degradations, leading to consistently strong ADD performance.

SILK is based on LPC and primarily optimized for voice communication.  It employs variable bitrate and bandwidth adaptation to cope with diverse network conditions while maintaining speech intelligibility. As can be observed from the second columns of Figure \ref{appenCodeclfccori}, \ref{appenCodeccqccori}, and \ref{appenCodecmfccori}, the t-SNE projections remain relatively stable across all TF features, with only minor shrinkage or deformation under high PLR. This stability indicates its effectiveness in preserving key TF spoofing features even under severe communication distortion, especially under the MFCC feature, leading to satisfactory ADD performance.

IVAS is still under development. The t-SNE visualization results of IVAS are presented in the third columns of Figure \ref{appenCodeclfccori}, \ref{appenCodeccqccori}, and \ref{appenCodecmfccori}. The t-SNE projections remain stable under MFCC and LFCC, with moderate structural deformation observed in CQCC from \(C_3\) to \(C_5\). Overall, IVAS retains sufficient spectral and temporal fidelity. This suggests that IVAS introduces relatively low distortion during encoding. Although loss and distortion of TF features occur under severe PLR, it still effectively preserves audio integrity under moderate degradation, leading to an acceptable ADD performance.

AMR-WB is based on Algebraic Code-Excited Linear Prediction (ACELP), a model designed to maintain speech intelligibility at low bitrates. 
ACELP may treat deepfake-specific anomalies as noise and aggressively suppress them through quantization or filtering. As PLR increases, its Packet Loss Concealment (PLC) mechanism relies more heavily on interpolation using typical speech patterns, which may oversmooth temporal and spectral variations. 
This behavior aligns with the projections of t-SNE shifts and structural distortions observed under high PLR, as shown in the fourth columns of Figure \ref{appenCodeclfccori}, \ref{appenCodeccqccori}, and \ref{appenCodecmfccori}. 
The observation indicates reasonable robustness and limited preservation of discriminative TF features, leading to a fair ADD performance.

Speex(WB) is based on Code-Excited Linear Prediction (CELP). As shown in the fifth columns of Figure \ref{appenCodeclfccori}, \ref{appenCodeccqccori}, and \ref{appenCodecmfccori}, a significant feature deformation occurs as PLR increases.
CELP tends to over-quantize or eliminate components that deviate from expected speech norms. If deepfake-specific features are identified as such anomalies, they are likely to be suppressed during encoding. Additionally, the PLC algorithm reconstructs missing frames based on conventional prior speech segments, further diminishing the presence of discriminative forgery artifacts. These effects reduce the codec’s ability to maintain ADD-relevant TF features under severe communication degradation, leading to a limited ADD performance.

EVS supports both ACELP and Modified Discrete Cosine Transform (MDCT)-based encoding modes depending on the bitrate and bandwidth. However, limitations of the ACELP in handling anomalous TF features still exist, including aggressive suppression of non-speech-like elements, PLC, and oversmoothing. As presented in the sixth columns of Figure \ref{appenCodeclfccori}, \ref{appenCodeccqccori}, and \ref{appenCodecmfccori}, the t-SNE projections exhibit significant structural distortions from \(C_2\) to \(C_5\), reflecting a substantial loss of discriminative features. These results suggest that EVS poses more challenges for ADD and leads to weaker detection performance compared to other speech codecs.

\begin{figure*}[h]
  \centering
  \includegraphics[width=0.8\textwidth]{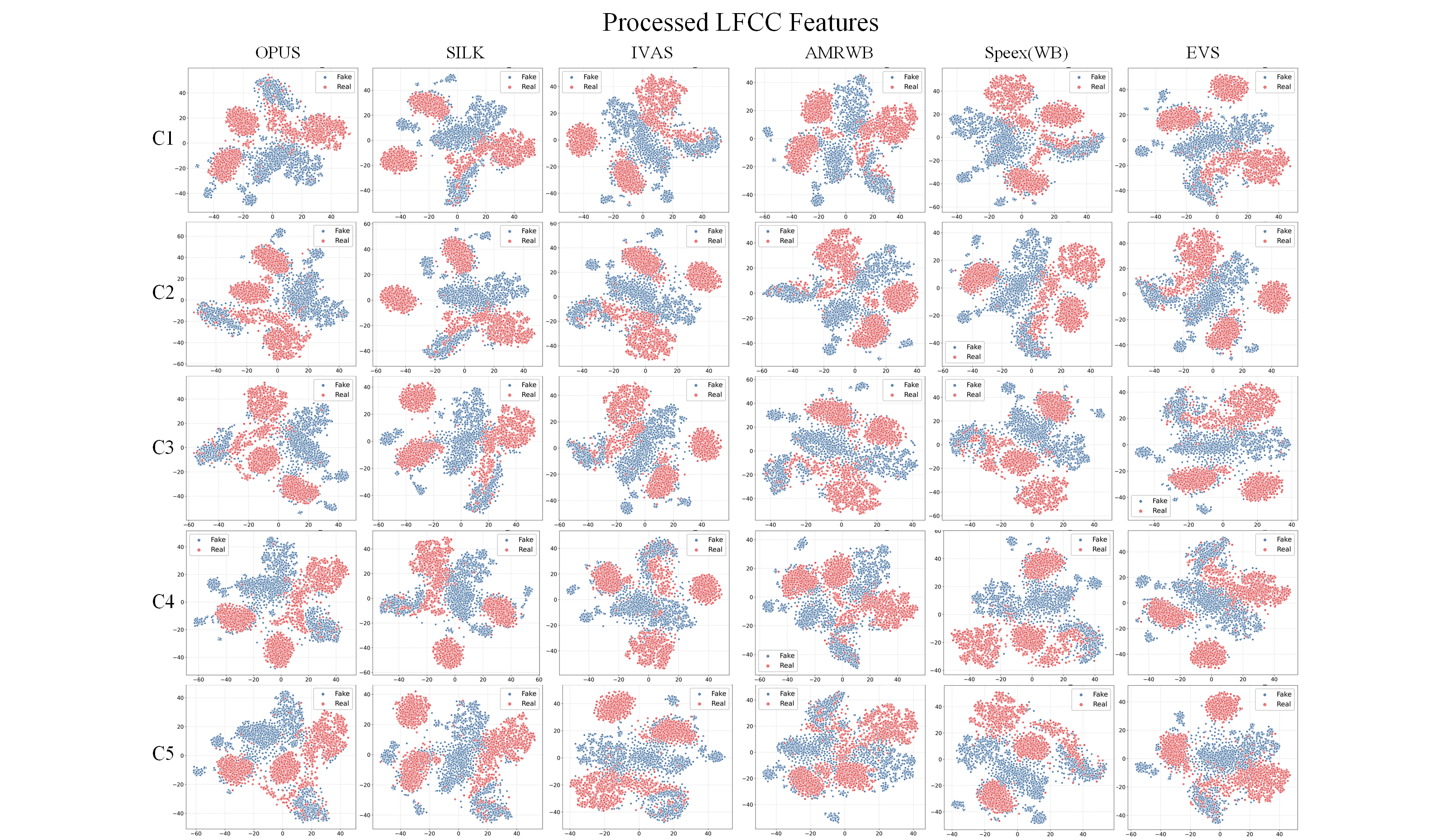}
  \caption{t-SNE visualizations of real and fake audio samples using the processed LFCC features, under six speech codecs (columns) and across \(C_1\) to \(C_5\) (rows).}
  \label{appenCodeclfccprocessed}
\end{figure*}

\begin{figure*}[h]
  \centering
  \includegraphics[width=0.8\textwidth]{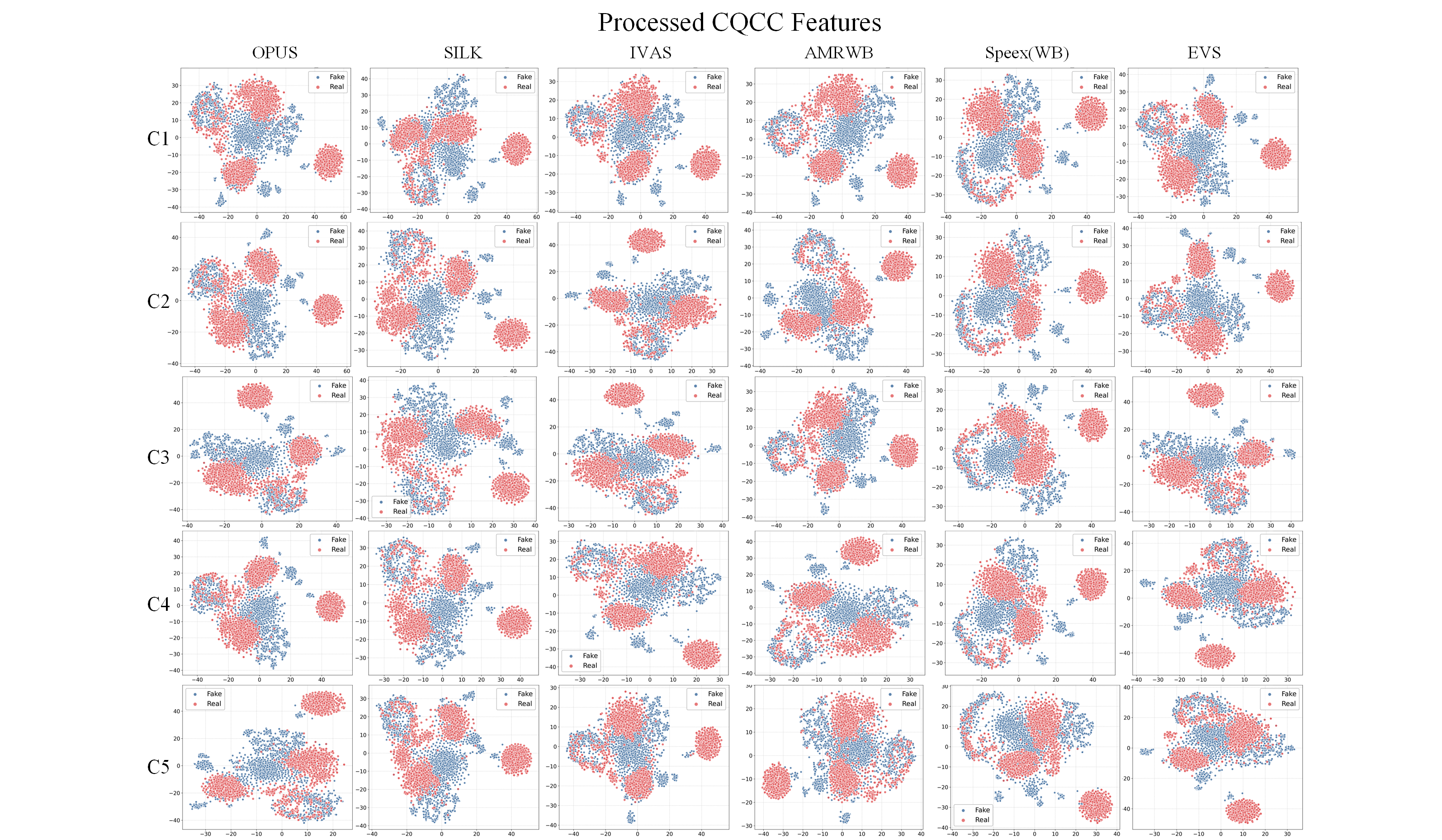}
  \caption{t-SNE visualizations of real and fake audio samples using the processed CQCC features, under six speech codecs (columns) and across \(C_1\) to \(C_5\) (rows).}
  \label{appenCodeccqccprocessed}
\end{figure*}

\begin{figure*}[h]
  \centering
  \includegraphics[width=0.8\textwidth]{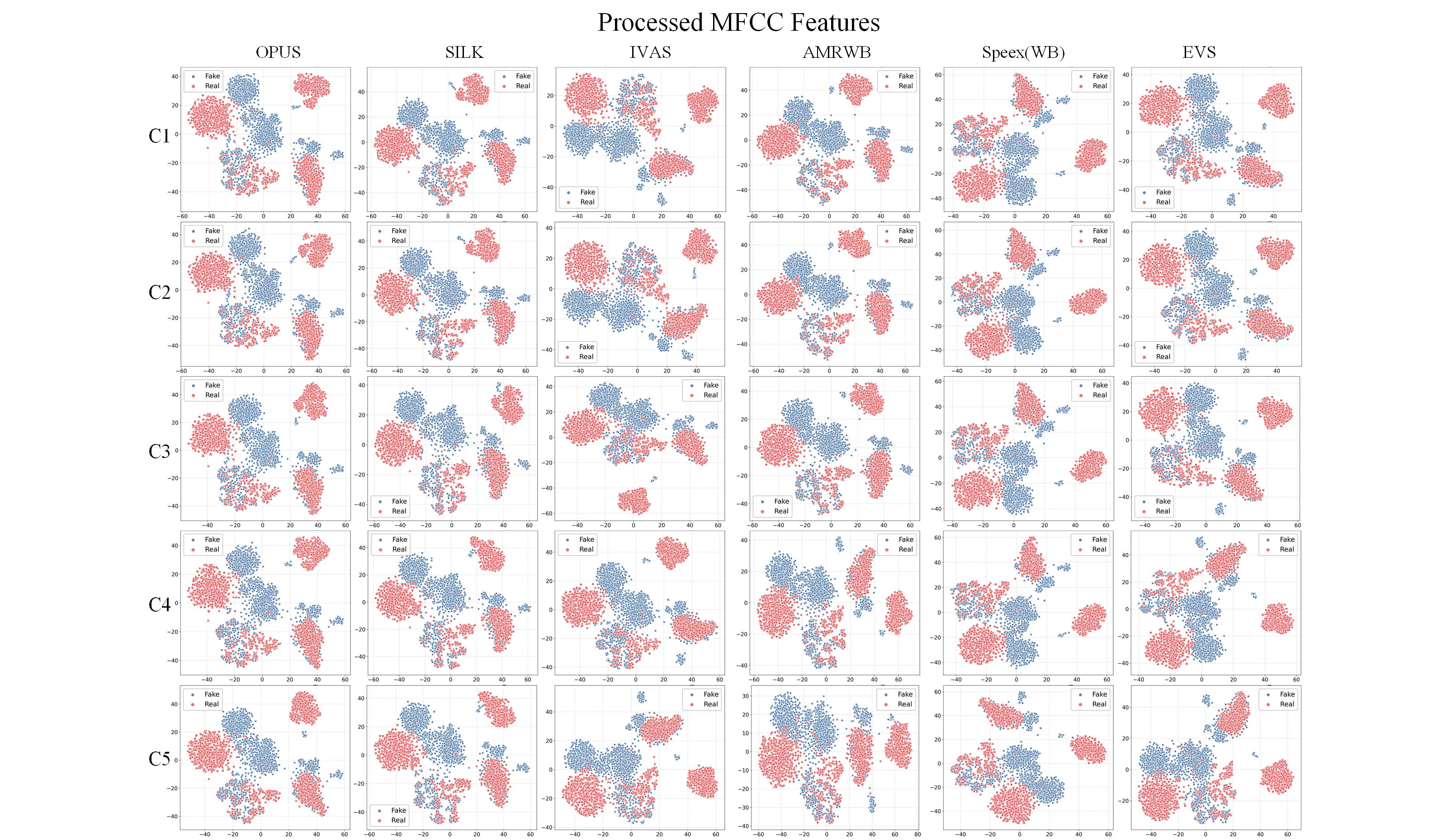}
  \caption{t-SNE visualizations of real and fake audio samples using the processed MFCC features, under six speech codecs (columns) and across \(C_1\) to \(C_5\) (rows).}
  \label{appenCodecmfccprocessed}
\end{figure*}

To further illustrate the effectiveness of the proposed framework, we provide the t-SNE visualizations for the processed TF features embeddings corresponding to Figure \ref{appenCodeclfccori}, \ref{appenCodeccqccori}, and \ref{appenCodecmfccori}, as shown in Figure \ref{appenCodeclfccprocessed}, \ref{appenCodeccqccprocessed}, and \ref{appenCodecmfccprocessed}, respectively. All features are extracted from the proposed framework before the Classifier. As can be observed, the processed embeddings exhibit well-formed clusters and clearly separated decision boundaries, confirming the framework’s robustness and effectiveness in improving and enhancing discriminative structures from the codec-specific aspect.


\section{Extended t-SNE visualization}\label{tsne}

\begin{figure*}[h]
  \centering
  \includegraphics[width=0.8\textwidth]{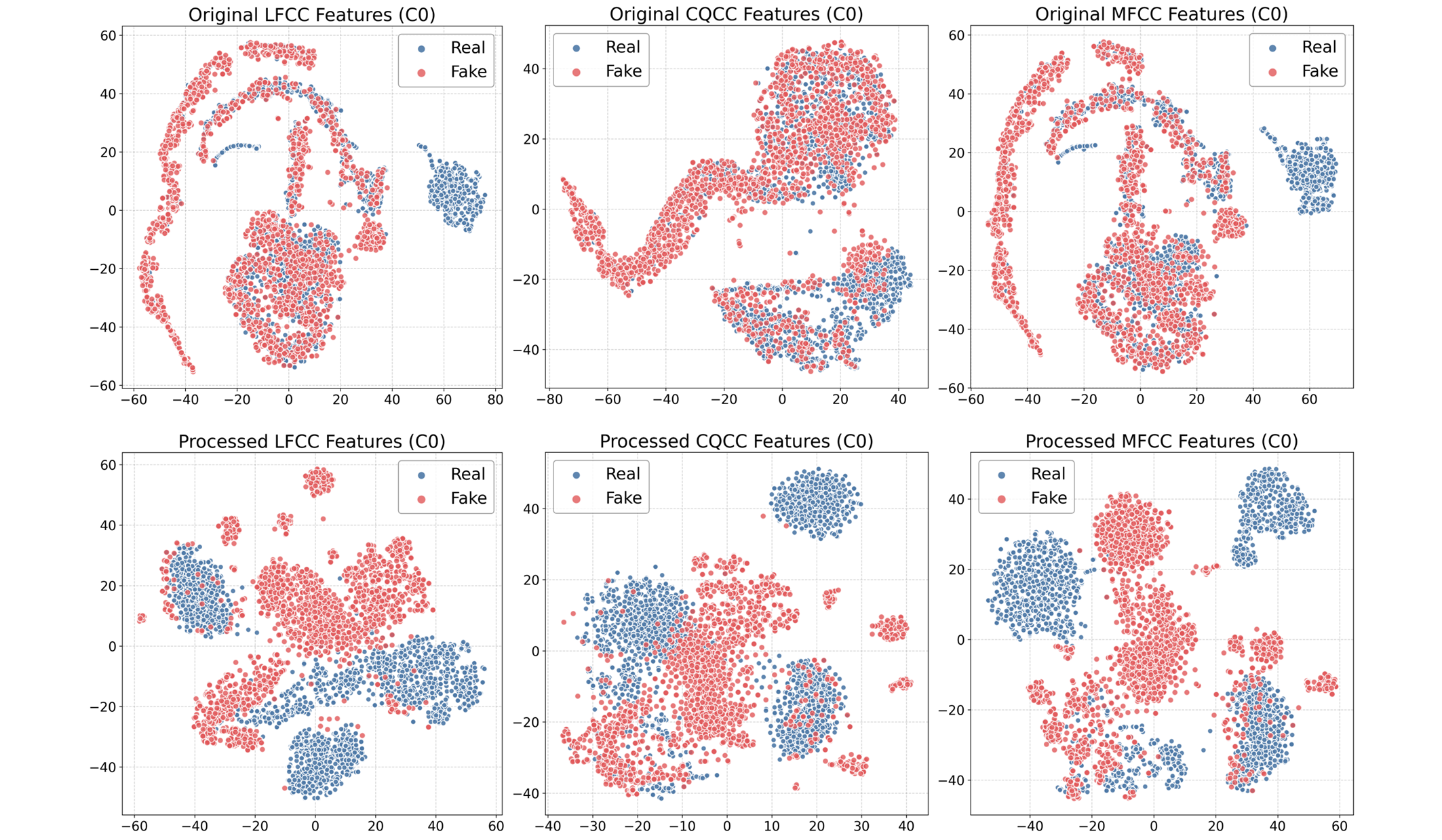}
  \caption{t-SNE visualizations of real and fake audio samples across different TF representations under \(C_0\). The top row represents the original features, and the bottom row represents the processed features extracted from the proposed framework before the Classifier.}
  \label{AppentsneC0}
\end{figure*}

\begin{figure*}[h]
  \centering
  \includegraphics[width=0.8\textwidth]{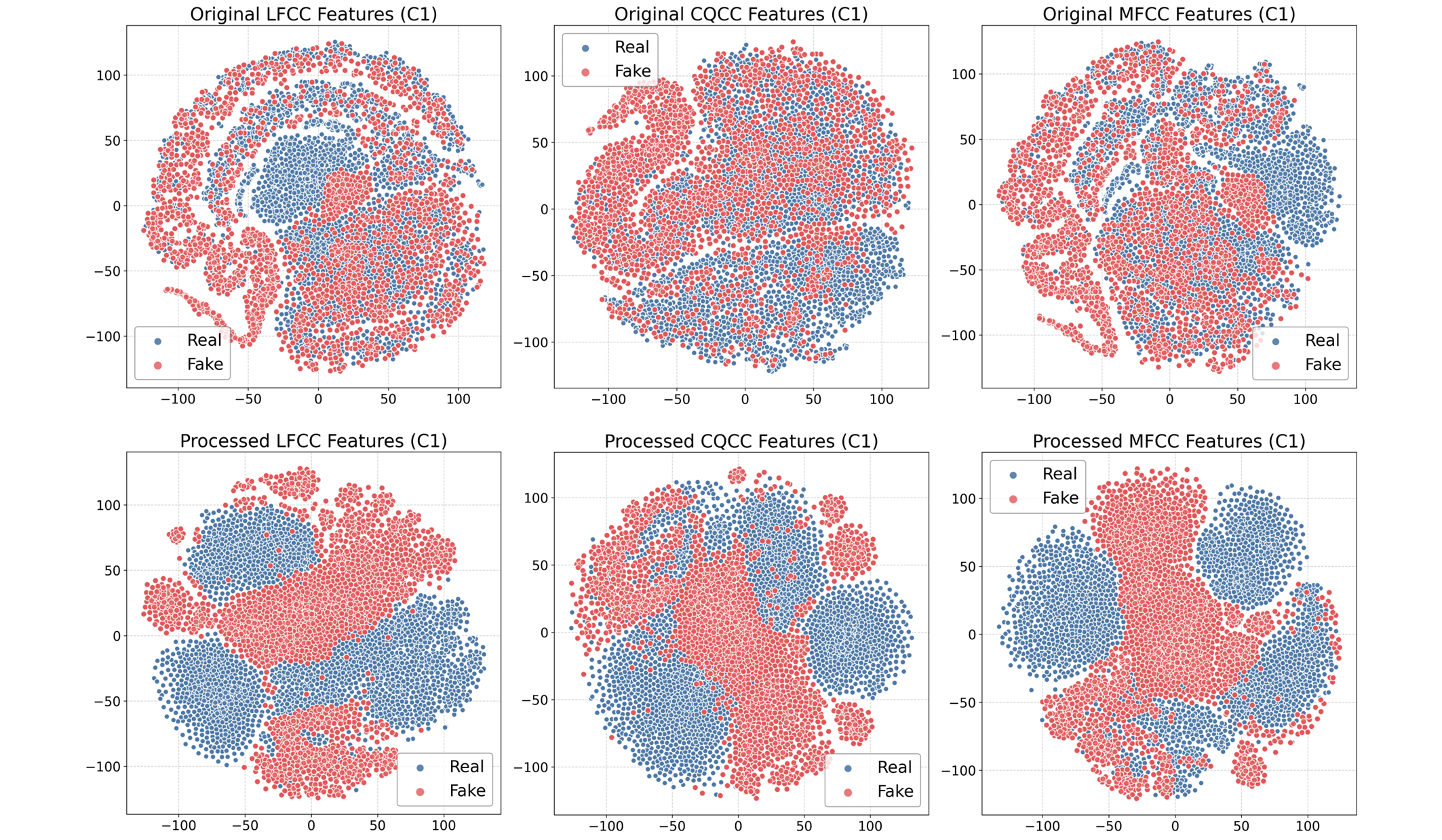}
  \caption{t-SNE visualizations of real and fake audio samples across different TF representations under \(C_1\). The top row represents the original features, and the bottom row represents the processed features extracted from the proposed framework before the Classifier.}
  \label{AppentsneC1}
\end{figure*}

\begin{figure*}[h]
  \centering
  \includegraphics[width=0.8\textwidth]{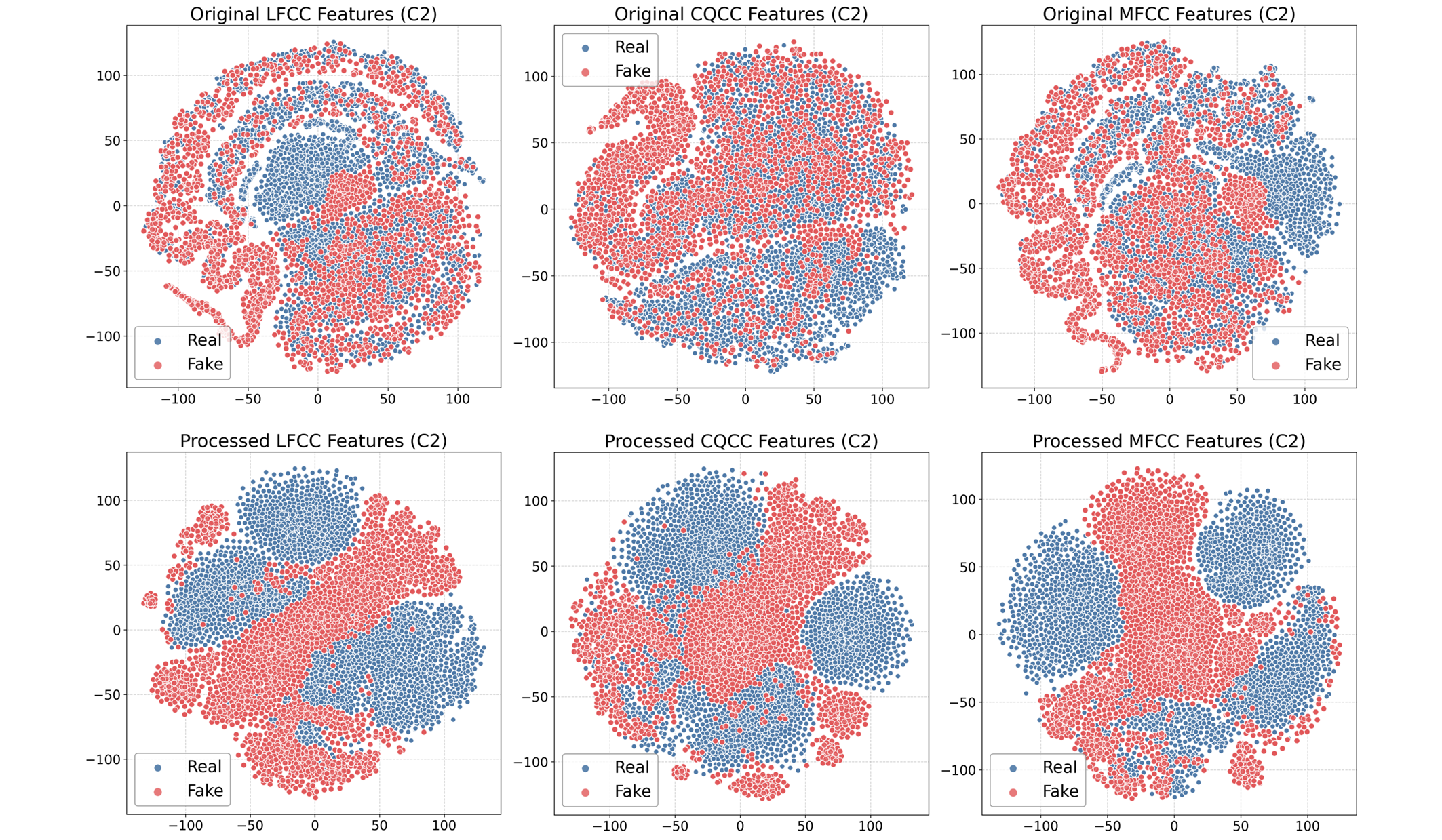}
  \caption{t-SNE visualizations of real and fake audio samples across different TF representations under \(C_2\). The top row represents the original features, and the bottom row represents the processed features extracted from the proposed framework before the Classifier.}
  \label{AppentsneC2}
\end{figure*}

\begin{figure*}[h]
  \centering
  \includegraphics[width=0.8\textwidth]{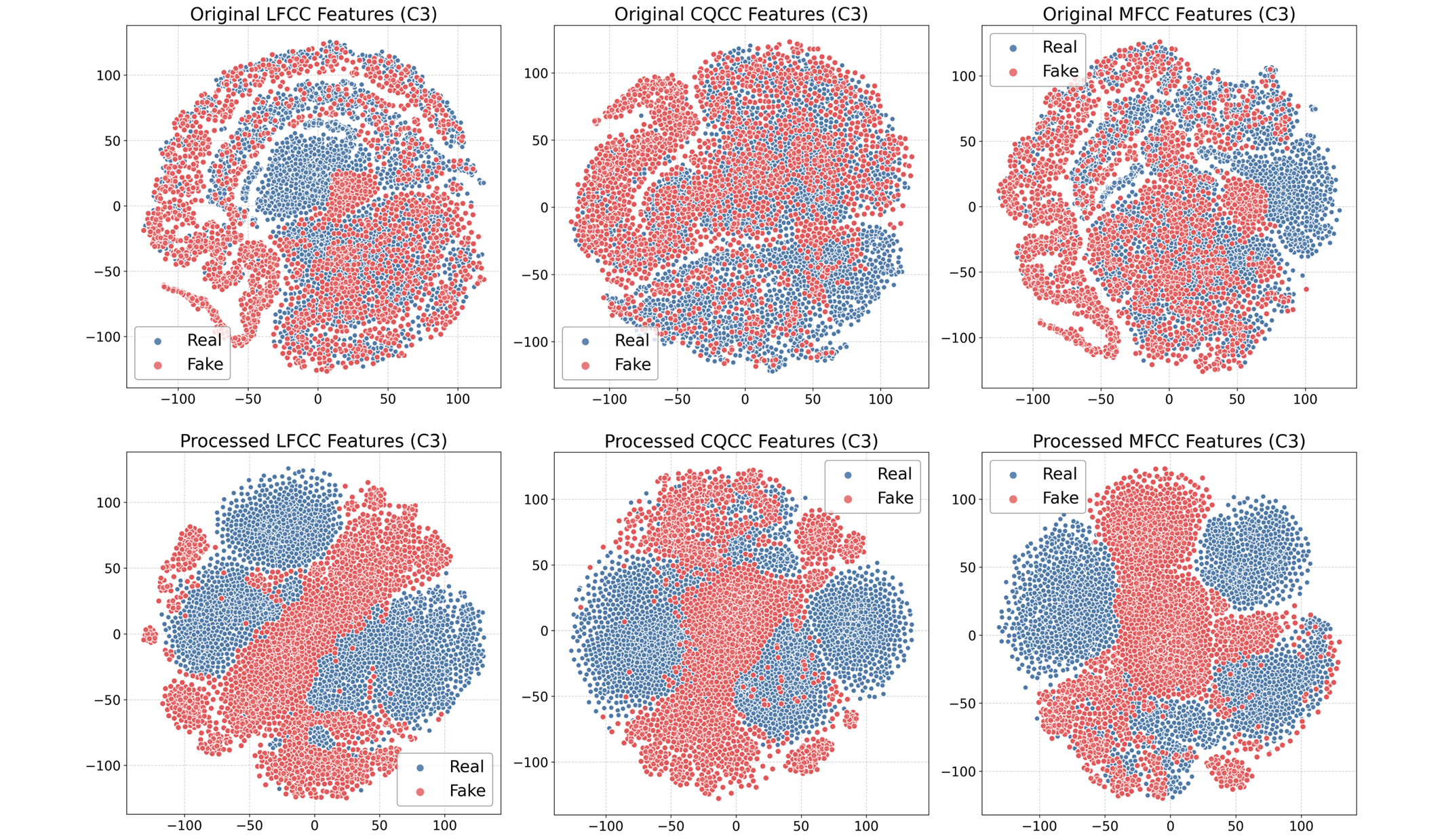}
  \caption{t-SNE visualizations of real and fake audio samples across different TF representations under \(C_3\). The top row represents the original features, and the bottom row represents the processed features extracted from the proposed framework before the Classifier.}
  \label{AppentsneC3}
\end{figure*}

\begin{figure*}[h]
  \centering
  \includegraphics[width=0.8\textwidth]{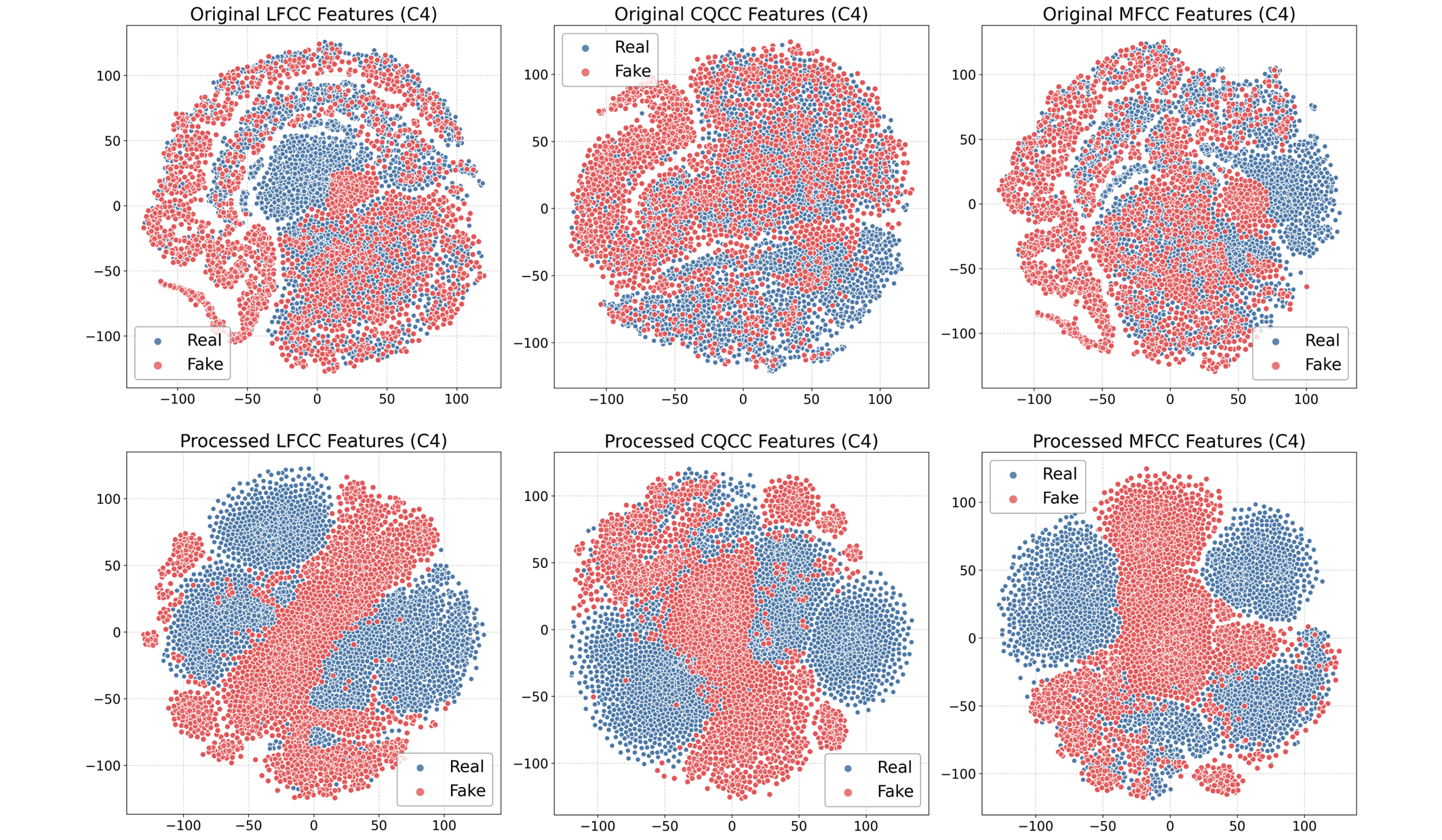}
  \caption{t-SNE visualizations of real and fake audio samples across different TF representations under \(C_4\). The top row represents the original features, and the bottom row represents the processed features extracted from the proposed framework before the Classifier.}
  \label{AppentsneC4}
\end{figure*}

To complement the analysis in Section \ref{tsnemain}, which presents the most severe case \(C_5\), the additional t-SNE visualizations from \(C_0\) to \(C_4\) are provided and shown in Figure \ref{AppentsneC0}, \ref{AppentsneC1}, \ref{AppentsneC2}, \ref{AppentsneC3} and \ref{AppentsneC4}, respectively.
Each figure illustrates real and fake audio samples across the three TF representations. 
These results offer a complete perspective on the framework's ability to effectively enhance class separability, not only under extreme degradations (\(C_5\)), but also across clean conditions (\(C_0\)) and increasingly severe real-world communication degradations (\(C_1\)-\(C_4\)). The consistently clear boundaries and reduced inter-class overlap further validate the proposed framework's effectiveness across a wide range of real-world communication degradations. 
Moreover, the results under the clean conditions further demonstrate its strong robustness and cross-domain generalization ability without requiring high-fidelity audio input, highlighting the practical deployment potential in real-world communication environments.

\end{document}